\documentclass[conference,compsoc]{IEEEtran}
\ifCLASSOPTIONcompsoc
  \usepackage[nocompress]{cite}
\else
  \usepackage{cite}
\fi
\usepackage{subcaption}
\usepackage{booktabs}
\usepackage{listings}
\ifCLASSINFOpdf
\else
\fi
  \usepackage{graphicx}
  \graphicspath{{./}{./figures/}{./data_import/figures/}}
  \usepackage{pgf}
\usepackage{array}
\usepackage{tabularx}
\usepackage{adjustbox}
\usepackage{rotating}
\usepackage{cuted}
\usepackage{pdflscape}

\usepackage{hyperref}
\usepackage{url}
\usepackage[table]{xcolor}
\usepackage{amsmath}
\usepackage{amsfonts}
\usepackage{amssymb}
\usepackage{tikz}
\usepackage{enumitem}
\usepackage[most]{tcolorbox}
\newtcolorbox{rqbox}{breakable, colback=gray!10, colframe=gray!40, boxrule=0.4pt, arc=0pt, left=6pt, right=6pt, top=4pt, bottom=4pt}  \newcommand{\rqtitle}

\definecolor{groupblue}{RGB}{232,240,254}
\definecolor{catgreen}{RGB}{218,239,203}
\definecolor{catred}{RGB}{250,219,216}

\usetikzlibrary{shapes.geometric, arrows.meta, positioning, fit, backgrounds}
\usepackage{pifont}
\let\oldding\ding
\newcommand\dingSize{1.2}
\newcommand\dingZero{171}
\renewcommand{\ding}[2][1]{\scalebox{#1}{\oldding{#2}}}
\newcommand{\circled}[1]{%
\raisebox{-0.3ex}{%
\ding[\dingSize]{\numexpr \dingZero+#1 \relax}}\xspace}

\newcommand{\Description}[1]{}

\usepackage[dvipsnames]{xcolor}
\usepackage{xspace}

\newcommand{\dyn}[1]{#1}
\newcommand{\nodyn}[1]{#1}
\renewcommand{\paragraph}[1]{\vspace{0.5em}\noindent\textbf{#1}.}
\newcommand{\essential}{essential\xspace}
\newcommand{\Essential}{Essential\xspace}
\newcommand{\critical}{sensitive\xspace}
\newcommand{\Critical}{Sensitive\xspace}
\newcommand{\discretion}{discretionary\xspace}
\newcommand{\Discretion}{Discretionary\xspace}
\newcommand{\expert}{baseline\xspace}

\newcommand{\baseline}{baseline\xspace}
\newcommand{\Baseline}{Baseline\xspace}

\begin{document}
\title{Can LLMs Make (Personalized) Access Control Decisions?}

\author{\IEEEauthorblockN{Friederike Groschupp\IEEEauthorrefmark{1},
Daniele Lain\IEEEauthorrefmark{1},
Aritra Dhar\IEEEauthorrefmark{2},
Lara Magdalena Lazier\IEEEauthorrefmark{2},
Srdjan \v{C}apkun\IEEEauthorrefmark{1}}
\IEEEauthorblockA{\IEEEauthorrefmark{1}Department of Computer Science, ETH Zurich, Switzerland}
\IEEEauthorblockN{\{friederike.groschupp, daniele.lain, srdjan.capkun\}@inf.ethz.ch}
\IEEEauthorblockA{\IEEEauthorrefmark{2}Computing Systems Lab, Huawei Technologies Switzerland AG}
\IEEEauthorblockN{\{aritra.dhar, lara.magdalena.lazier2\}@huawei.com}
}

\maketitle
\pagestyle{plain}

\begin{abstract}
Precise access control decisions are crucial for the security of both traditional applications and emerging agent-based systems. Typically, these decisions are made by users during app installation or at runtime. However, due to the increasing complexity and automation of systems, making access control decisions can impose a significant cognitive burden on users, often overwhelming them and leading to suboptimal or even arbitrary choices.

To address this problem, we investigate the ability of LLMs to make dynamic, context-aware decisions aligned with users' security preferences, expressed during a lightweight setup phase. As a case study, we analyze smartphone application permission requests, given their ubiquity and users’ familiarity with them. We curated a dataset comprising 307 user \textit{privacy statements} (short, natural-language descriptions of user preferences) and 14,682 corresponding permission decisions, gathered from smartphone users in an online data collection. We compare these decisions with those made by two versions of LLMs that are tasked with reasoning about the app and the request context: a general model and a personalized one (which incorporates user preferences). For the latter, we also collected user feedback on 1,298 of its decisions.

Our results show that LLMs generally reflect users’ preferences well, agreeing with the majority decision in up to 86\% of cases, and can steer users toward safer behavior. However, the results also reveal a key trade-off in personalization: while incorporating user-specific privacy preferences improves agreement with individual decisions, strict adherence to these preferences may lead to less safe outcomes, as users tend to over-permission.

\end{abstract}

\IEEEpeerreviewmaketitle

\newcommand{\NumProlific}{393}
\newcommand{\NumValidProlific}{307}
\newcommand{\NumOptional}{184}
\newcommand{\MedAge}{42.00}
\newcommand{\StdAge}{12.67}
\newcommand{\AvgPermKnow}{3.45}
\newcommand{\StdPermKnow}{1.01}
\newcommand{\MedTimeMins}{32.58}
\newcommand{\StdTimeMins}{16.40}
\newcommand{\PctFemale}{50.81}
\newcommand{\PctMale}{48.53}
\newcommand{\PctOtherGender}{0.33}
\newcommand{\PctAndroid}{88.93}
\newcommand{\PctiOS}{11.07}
\newcommand{\PctUK}{49.19}
\newcommand{\PctUS}{50.81}
\newcommand{\PctEnglish}{93.49}

\newcommand{\gptSMajorityAllAccuracy}{78}
\newcommand{\gptSMajorityWoutAccuracy}{80}
\newcommand{\gptSMajorityCatScenarioExpertAllowAccuracy}{100}
\newcommand{\gptSMajorityCatScenarioExpertDenyAccuracy}{67}
\newcommand{\gptSMajorityCatScenarioNoExpertAccuracy}{67}
\newcommand{\gptAIMajorityAllPearsonR}{-0.02}
\newcommand{\gptAIMajorityAllPearsonDF}{109}
\newcommand{\gptAIMajorityAllAccuracy}{86}
\newcommand{\gptAIMajorityWoutAccuracy}{89}
\newcommand{\gptAIMajorityCatScenarioExpertAllowAccuracy}{100}
\newcommand{\gptAIMajorityCatScenarioExpertDenyAccuracy}{67}
\newcommand{\gptAIMajorityCatScenarioNoExpertAccuracy}{73}
\newcommand{\gptAoMajorityAllPearsonR}{0.51}
\newcommand{\gptAoMajorityAllPearsonDF}{109}
\newcommand{\gptAoMajorityAllAccuracy}{80}
\newcommand{\gptAoMajorityWoutPearsonR}{0.50}
\newcommand{\gptAoMajorityWoutPearsonDF}{82}
\newcommand{\gptAoMajorityWoutAccuracy}{84}
\newcommand{\gptAoMajorityWithPearsonR}{0.54}
\newcommand{\gptAoMajorityWithPearsonDF}{25}
\newcommand{\gptAoMajorityCatScenarioExpertAllowAccuracy}{100}
\newcommand{\gptAoMajorityCatScenarioExpertDenyAccuracy}{50}
\newcommand{\gptAoMajorityCatScenarioNoExpertAccuracy}{60}
\newcommand{\mistralMajorityAllAccuracy}{84}
\newcommand{\mistralMajorityWoutAccuracy}{82}
\newcommand{\mistralMajorityCatScenarioExpertAllowAccuracy}{100}
\newcommand{\mistralMajorityCatScenarioExpertDenyAccuracy}{83}
\newcommand{\mistralMajorityCatScenarioNoExpertAccuracy}{87}
\newcommand{\grokMajorityAllAccuracy}{85}
\newcommand{\grokMajorityWoutAccuracy}{88}
\newcommand{\grokMajorityCatScenarioExpertAllowAccuracy}{100}
\newcommand{\grokMajorityCatScenarioExpertDenyAccuracy}{83}
\newcommand{\grokMajorityCatScenarioNoExpertAccuracy}{67}
\newcommand{\deepseekMajorityAllAccuracy}{75}
\newcommand{\deepseekMajorityWoutAccuracy}{75}
\newcommand{\deepseekMajorityCatScenarioExpertAllowAccuracy}{100}
\newcommand{\deepseekMajorityCatScenarioExpertDenyAccuracy}{50}
\newcommand{\deepseekMajorityCatScenarioNoExpertAccuracy}{73}

\newcommand{\personalizedPerUserMedianAcc}{71.74}
\newcommand{\personalizedPerUserStdAcc}{13.88}
\newcommand{\personalizedPerUserMinAcc}{21.95}
\newcommand{\personalizedPerUserMaxAcc}{100}
\newcommand{\genericPerUserMedianAcc}{67.31}
\newcommand{\genericPerUserStdAcc}{10.16}
\newcommand{\genericPerUserMinAcc}{38.46}
\newcommand{\genericPerUserMaxAcc}{89.80}
\newcommand{\mistralPersonalizedPerUserMedianAcc}{72.00}
\newcommand{\mistralPersonalizedPerUserStdAcc}{15.76}
\newcommand{\mistralPersonalizedPerUserMinAcc}{17.07}
\newcommand{\mistralPersonalizedPerUserMaxAcc}{100}
\newcommand{\mistralGenericPerUserMedianAcc}{70.73}
\newcommand{\mistralGenericPerUserStdAcc}{14.14}
\newcommand{\mistralGenericPerUserMinAcc}{26.92}
\newcommand{\mistralGenericPerUserMaxAcc}{91.49}
\newcommand{\diffPerUserDF}{306}
\newcommand{\diffPerUserMean}{4.02}
\newcommand{\diffPerUserMin}{-33.33}
\newcommand{\diffPerUserMax}{54.55}
\newcommand{\pairedTTestT}{5.976}
\newcommand{\mistralDiffPerUserDF}{306}
\newcommand{\mistralDiffPerUserMean}{2.56}
\newcommand{\mistralPairedTTestT}{4.544}

\newcommand{\OverallTotal}{14,682}
\newcommand{\OverallDenyRate}{56}
\newcommand{\MinDecisionsPerTask}{90}
\newcommand{\MaxDecisionsPerTask}{209}
\newcommand{\DenyRateNoScenario}{60}
\newcommand{\DenyRateExpertAllow}{8}
\newcommand{\DenyRateExpertDeny}{55}
\newcommand{\DenyRateNoExpert}{54}
\newcommand{\ClearMajNoScenarioCount}{38}
\newcommand{\ClearMajNoScenarioTotal}{84}
\newcommand{\ClearMajWithNoExpertCount}{4}
\newcommand{\ClearMajWithNoExpertTotal}{15}
\newcommand{\ClearMajorityThreshold}{75}
\newcommand{\CategoriesExamplesNoScenario}{84}
\newcommand{\CategoriesExamplesScenarioExpertAllow}{6}
\newcommand{\CategoriesExamplesScenarioExpertDeny}{6}
\newcommand{\CategoriesExamplesScenarioNoExpert}{15}
\newcommand{\CategoriesDecisionsNoScenario}{10,472}
\newcommand{\CategoriesDecisionsScenarioExpertAllow}{921}
\newcommand{\CategoriesDecisionsScenarioExpertDeny}{981}
\newcommand{\CategoriesDecisionsScenarioNoExpert}{2,308}

\newcommand{\gptAoVsUsersOverallAcc}{66.49}
\newcommand{\gptAoVsUsersOverallViolations}{16.36}
\newcommand{\gptAoVsUsersOverallOverRestrictions}{17.15}
\newcommand{\gptAoVsUsersAccNoScenario}{67.78}
\newcommand{\gptAoVsUsersViolationsNoScenario}{13.12}
\newcommand{\gptAoVsUsersOverRestrictionsNoScenario}{19.10}
\newcommand{\gptAoVsUsersAccScenarioExpertAllow}{92.18}
\newcommand{\gptAoVsUsersViolationsScenarioExpertAllow}{7.82}
\newcommand{\gptAoVsUsersOverRestrictionsScenarioExpertAllow}{0.00}
\newcommand{\gptAoVsUsersAccScenarioExpertDeny}{50.97}
\newcommand{\gptAoVsUsersViolationsScenarioExpertDeny}{10.19}
\newcommand{\gptAoVsUsersOverRestrictionsScenarioExpertDeny}{38.84}
\newcommand{\gptAoVsUsersAccScenarioNoExpert}{56.98}
\newcommand{\gptAoVsUsersViolationsScenarioNoExpert}{37.09}
\newcommand{\gptAoVsUsersOverRestrictionsScenarioNoExpert}{5.94}
\newcommand{\mistralVsUsersOverallAcc}{67.40}
\newcommand{\mistralVsUsersOverallViolations}{10.09}
\newcommand{\mistralVsUsersOverallOverRestrictions}{22.52}
\newcommand{\mistralVsUsersAccNoScenario}{66.30}
\newcommand{\mistralVsUsersViolationsNoScenario}{8.72}
\newcommand{\mistralVsUsersOverRestrictionsNoScenario}{24.98}
\newcommand{\mistralVsUsersAccScenarioExpertAllow}{92.18}
\newcommand{\mistralVsUsersViolationsScenarioExpertAllow}{7.82}
\newcommand{\mistralVsUsersOverRestrictionsScenarioExpertAllow}{0.00}
\newcommand{\mistralVsUsersAccScenarioExpertDeny}{61.06}
\newcommand{\mistralVsUsersViolationsScenarioExpertDeny}{4.38}
\newcommand{\mistralVsUsersOverRestrictionsScenarioExpertDeny}{34.56}
\newcommand{\mistralVsUsersAccScenarioNoExpert}{65.16}
\newcommand{\mistralVsUsersViolationsScenarioNoExpert}{19.63}
\newcommand{\mistralVsUsersOverRestrictionsScenarioNoExpert}{15.21}

\newcommand{\gptAoPersonalizedVsUsersOverallAcc}{70.42}
\newcommand{\gptAoPersonalizedVsUsersOverallViolations}{10.43}
\newcommand{\gptAoPersonalizedVsUsersOverallOverRestrictions}{19.15}
\newcommand{\mistralPersonalizedVsUsersOverallAcc}{69.93}
\newcommand{\mistralPersonalizedVsUsersOverallViolations}{6.42}
\newcommand{\mistralPersonalizedVsUsersOverallOverRestrictions}{23.65}
\newcommand{\gptAoPersonalizedVsUsersAccNoScenario}{70.98}
\newcommand{\gptAoPersonalizedVsUsersViolationsNoScenario}{9.34}
\newcommand{\gptAoPersonalizedVsUsersOverRestrictionsNoScenario}{19.68}
\newcommand{\gptAoPersonalizedVsUsersAccScenarioExpertAllow}{91.42}
\newcommand{\gptAoPersonalizedVsUsersViolationsScenarioExpertAllow}{7.60}
\newcommand{\gptAoPersonalizedVsUsersOverRestrictionsScenarioExpertAllow}{0.98}
\newcommand{\gptAoPersonalizedVsUsersAccScenarioExpertDeny}{55.25}
\newcommand{\gptAoPersonalizedVsUsersViolationsScenarioExpertDeny}{6.52}
\newcommand{\gptAoPersonalizedVsUsersOverRestrictionsScenarioExpertDeny}{38.23}
\newcommand{\gptAoPersonalizedVsUsersAccScenarioNoExpert}{65.94}
\newcommand{\gptAoPersonalizedVsUsersViolationsScenarioNoExpert}{18.20}
\newcommand{\gptAoPersonalizedVsUsersOverRestrictionsScenarioNoExpert}{15.86}
\newcommand{\mistralPersonalizedVsUsersAccNoScenario}{69.83}
\newcommand{\mistralPersonalizedVsUsersViolationsNoScenario}{4.69}
\newcommand{\mistralPersonalizedVsUsersOverRestrictionsNoScenario}{25.48}
\newcommand{\mistralPersonalizedVsUsersAccScenarioExpertAllow}{90.99}
\newcommand{\mistralPersonalizedVsUsersViolationsScenarioExpertAllow}{6.84}
\newcommand{\mistralPersonalizedVsUsersOverRestrictionsScenarioExpertAllow}{2.17}
\newcommand{\mistralPersonalizedVsUsersAccScenarioExpertDeny}{61.57}
\newcommand{\mistralPersonalizedVsUsersViolationsScenarioExpertDeny}{8.77}
\newcommand{\mistralPersonalizedVsUsersOverRestrictionsScenarioExpertDeny}{29.66}
\newcommand{\mistralPersonalizedVsUsersAccScenarioNoExpert}{65.51}
\newcommand{\mistralPersonalizedVsUsersViolationsScenarioNoExpert}{13.08}
\newcommand{\mistralPersonalizedVsUsersOverRestrictionsScenarioNoExpert}{21.40}

\newcommand{\MismatchExpertPersTotal}{518}
\newcommand{\MismatchExpertPersAccuracy}{86}
\newcommand{\MismatchExpertPersTotalAllow}{79}
\newcommand{\MismatchExpertPersTotalDeny}{439}
\newcommand{\MismatchExpertPersAccuracyAllow}{89}
\newcommand{\MismatchExpertPersAccuracyDeny}{85}

\newcommand{\PersInputLenCorrPearsonR}{0.183}
\newcommand{\PersInputLenCorrDF}{305}

\newcommand{\AnovaGroupF}{0.072}
\newcommand{\AnovaGroupP}{0.7890}
\newcommand{\AnovaGroupDF}{1}
\newcommand{\AnovaStyleF}{0.517}
\newcommand{\AnovaStyleP}{0.4728}
\newcommand{\AnovaStyleDF}{1}
\newcommand{\AnovaInteractionF}{4.328}
\newcommand{\AnovaInteractionP}{0.0383}
\newcommand{\AnovaInteractionDF}{1}
\newcommand{\AnovaResidualDF}{303}

\newcommand{\PhoneCfgCompareTotal}{472}
\newcommand{\PhoneCfgCompareAccuracy}{77.33}
\newcommand{\PhoneCfgCameraAccuracy}{63.83}
\newcommand{\PhoneCfgMicrophoneAccuracy}{70.51}
\newcommand{\PhoneCfgPhotosAccuracy}{86.96}
\newcommand{\PhoneCfgCalendarAccuracy}{87.50}

\newcommand{\LlmRateYesInitialagreed}{99.44}
\newcommand{\LlmRateNoInitialagreed}{0.37}
\newcommand{\LlmRateNotSureInitialagreed}{0.19}
\newcommand{\LlmRateYesInitialdisagreed}{48.61}
\newcommand{\LlmRateNoInitialdisagreed}{42.06}
\newcommand{\LlmRateNotSureInitialdisagreed}{9.33}
\newcommand{\LlmRateYesInitialallowvsonce}{77.03}
\newcommand{\LlmRateNoInitialallowvsonce}{18.92}
\newcommand{\LlmRateNotSureInitialallowvsonce}{4.05}
\newcommand{\LlmOverallYesRate}{72.96}
\newcommand{\LlmOverallNoRate}{22.11}
\newcommand{\LlmOverallNotSureRate}{4.93}
\newcommand{\LlmMismatchYesRateUserAllow}{59.82}
\newcommand{\LlmMismatchNoRateUserAllow}{30.67}
\newcommand{\LlmMismatchNotSureRateUserAllow}{9.51}
\newcommand{\LlmMismatchYesRateUserDeny}{27.27}
\newcommand{\LlmMismatchNoRateUserDeny}{64.09}
\newcommand{\LlmMismatchNotSureRateUserDeny}{8.64}
\newcommand{\LlmMismatchYesRateUserOnce}{64.62}
\newcommand{\LlmMismatchNoRateUserOnce}{24.62}
\newcommand{\LlmMismatchNotSureRateUserOnce}{10.77}
\newcommand{\LlmMismatchOverallYesRate}{48.61}
\newcommand{\LlmMismatchOverallNoRate}{42.06}
\newcommand{\LlmMismatchOverallNotSureRate}{9.33}
\newcommand{\LlmComboUserAllowLlmAllowYesPersonalPct}{38}
\newcommand{\LlmComboUserAllowLlmAllowNoPersonalPct}{0}
\newcommand{\LlmComboUserAllowLlmAllowYesDetailsPct}{25}
\newcommand{\LlmComboUserAllowLlmAllowYesAppPct}{38}
\newcommand{\LlmComboUserAllowLlmAllowYesOtherPct}{0}
\newcommand{\LlmComboUserAllowLlmAllowNoOtherPct}{0}
\newcommand{\LlmComboUserAllowLlmDenyYesPersonalPct}{37}
\newcommand{\LlmComboUserAllowLlmDenyNoPersonalPct}{32}
\newcommand{\LlmComboUserAllowLlmDenyYesDetailsPct}{26}
\newcommand{\LlmComboUserAllowLlmDenyNoDetailsPct}{17}
\newcommand{\LlmComboUserAllowLlmDenyYesAppPct}{34}
\newcommand{\LlmComboUserAllowLlmDenyNoAppPct}{33}
\newcommand{\LlmComboUserAllowLlmDenyYesOtherPct}{3}
\newcommand{\LlmComboUserAllowLlmDenyNoOtherPct}{18}
\newcommand{\LlmComboUserDenyLlmAllowYesPersonalPct}{22}
\newcommand{\LlmComboUserDenyLlmAllowNoPersonalPct}{54}
\newcommand{\LlmComboUserDenyLlmAllowYesDetailsPct}{31}
\newcommand{\LlmComboUserDenyLlmAllowNoDetailsPct}{8}
\newcommand{\LlmComboUserDenyLlmAllowYesAppPct}{46}
\newcommand{\LlmComboUserDenyLlmAllowNoAppPct}{23}
\newcommand{\LlmComboUserDenyLlmAllowYesOtherPct}{2}
\newcommand{\LlmComboUserDenyLlmAllowNoOtherPct}{15}
\newcommand{\LlmComboUserDenyLlmDenyYesPersonalPct}{44}
\newcommand{\LlmComboUserDenyLlmDenyNoPersonalPct}{0}
\newcommand{\LlmComboUserDenyLlmDenyYesDetailsPct}{22}
\newcommand{\LlmComboUserDenyLlmDenyNoDetailsPct}{50}
\newcommand{\LlmComboUserDenyLlmDenyYesAppPct}{34}
\newcommand{\LlmComboUserDenyLlmDenyNoAppPct}{50}
\newcommand{\LlmComboUserDenyLlmDenyYesOtherPct}{0}
\newcommand{\LlmComboUserDenyLlmDenyNoOtherPct}{0}
\newcommand{\LlmComboTotalYesPersonalPct}{40}
\newcommand{\LlmComboTotalNoPersonalPct}{42}
\newcommand{\LlmComboTotalYesDetailsPct}{24}
\newcommand{\LlmComboTotalNoDetailsPct}{13}
\newcommand{\LlmComboTotalYesAppPct}{36}
\newcommand{\LlmComboTotalNoAppPct}{28}
\newcommand{\LlmComboTotalYesOtherPct}{1}
\newcommand{\LlmComboTotalNoOtherPct}{17}

\newcommand{\PersonalizedVsExpertAccuracyAllow}{98.81}
\newcommand{\PersonalizedVsExpertAccuracyDeny}{86.51}

\newcommand{\USOverallDenyRate}{55}
\newcommand{\UKOverallDenyRate}{56}
\newcommand{\CountryDenyRateTestZ}{-1.73}
\newcommand{\CountryDenyRateTestP}{.083}

\newcommand{\USPerUserNUsers}{156}
\newcommand{\UKPerUserNUsers}{151}
\newcommand{\CountryPerUserTestT}{-1.58}
\newcommand{\CountryPerUserTestP}{.116}

\section{Introduction}

With the rising capabilities of large language models (LLMs), they are increasingly employed to solve classical security problems, such as synthesizing firewall rules~\cite{huang2024large}, detecting security bugs~\cite{mhatre2025llm}, or formal verification~\cite{orenes2023using}, among others. 
We turn our attention to access control, an area known for imposing a high cognitive burden on users~\cite{felt2012android} that often results in suboptimal decisions~\cite{egelman2013choice,fang2010privacy}. 

We investigate whether we can reduce this burden by relying on state-of-the-art LLMs to make personalized runtime decisions on behalf of the user, specifically in the context of smartphone app permissions. 
To reduce the setup cost of such a system, we want to avoid approaches that would require gathering a large set of data points from the user or require retraining or fine-tuning the model to learn their preferences. 
Instead, we ask the user to provide a \textit{privacy statement}, a short natural-language description of their access control preferences. The LLM then makes decisions based on this statement, the request context, and its trained knowledge, balancing user requirements and security considerations.

Using LLMs to make safe access control decisions that reflect users' privacy preferences is non-trivial and poses several challenges.
First, the quality of LLM inference is dependent on the input~\cite{he2024doespromptformattingimpact}. 
Therefore, poorly expressed privacy preferences may result in less accurate access control decisions.
Second, it is well documented that self-reported behavior, particularly in the context of security, often suffers from bias~\cite{fisher2000social} and the privacy paradox~\cite{norberg2007privacy,acquisti2015privacy}, resulting in a divergence between reported preference and actual actions~\cite{wash2010folk}.
Third, limitations of LLMs, such as hallucinations~\cite{rawte2023survey} or biases~\cite{bias_llms}, might lead to undesired decision-making.
In this work, we therefore ask the following question:
\textit{Can LLMs derive users' privacy preferences from a few user-provided instructions and make access control decisions on their behalf?}

To answer this question, we collected a dataset in the context of smartphone permission management, comprising users' privacy preferences in natural language, their decisions for different apps requesting access to different resources, some with described scenarios, and participants' feedback on  LLM decisions. To the best of our knowledge, this is the first dataset comprising both personal natural-language privacy statements and
corresponding access control decisions.
This dataset allows us to systematically evaluate LLM access control decision-making in terms of agreement with human consensus, individual user preferences, and safety.

Our results show that LLMs can indeed reflect general human consensus,
matching the decision preferred by the majority of participants for up to 86\% of our tasks and mostly agreeing with standard practices. 
However, comparing to individuals shows a significant variation in agreement, with some users having as low as 27\% agreement. 
Personalization based on user-provided privacy statements improves agreement, but results vary significantly across users: some see significant improvements (up to 100\% agreement), while others experience a decrease in agreement. The success of personalization appears to be linked to the quality of the privacy statement provided by users and its alignment with their decisions, as well as to the extent to which their preferences deviate from the general population (e.g., very privacy-conscious users benefit disproportionately from personalization). 

However, agreement with the initial user decision is not always desirable: when the user and the LLM initially disagreed, follow-up feedback revealed that users actually preferred the LLM choice 49\% of the time.
Furthermore, on tasks where we defined an expected outcome (e.g., access requests clearly unrelated to the purpose of an app should be denied), the decision of the LLM is often safer than the user's decision. While there are some instances where the LLM makes wrong decisions both compared to user preferences and our expected baseline, we estimate the overall LLM success rate to be between 79\% and 99\% across different task types.
Our results indicate that LLM-supported smartphone permission management can be a viable tool to alleviate user burden and support security-aware behavior, but its deployment needs careful design. They also warrant further investigation into whether this generalizes to other types of access control systems.

\paragraph{Contributions}
In summary, our contributions are:
\begin{itemize}[nosep]
    \item We measure the extent to which LLMs make ``reasonable'' access control decisions for smartphone app permissions,  consistent with expected safe behaviors, common user choices, and personal preferences, by collecting and analyzing a new dataset of 307 privacy statements and 14,682 user and LLM access control decisions.
    \item We show that LLMs can make secure, informed permission decisions based on task context and their system knowledge, even steering human judgment when users make mistakes. Personalization can greatly increase agreement for some users, but its usage requires caution.
    
    \item We discuss setup considerations for implementing a practical access control system that balances personalization, security, and utility by deferring uncertain decisions to the user.
\end{itemize}

\section{Motivation and Methodology}
\label{sec:motivation}

Making access control decisions is known to be a significant cognitive burden~\cite{felt2012android} for users. 
Factors such as lack of technical knowledge, insufficient attention, or time pressure can lead to decisions that are inconsistent with the user's internal beliefs~\cite{vance2019} and to data leaks or other irreversible consequences~\cite{felt2011android,wijesekera2015android,egelman2013choice,wijesekera2017feasibility,fang2010privacy}. 
Consequently, previous work has explored automating such decisions. Traditional machine-learning approaches~\cite{wijesekera2018contextualizing,  10.1145/3508398.3511497, turtleguard, DBLP:conf/soups/0017ASAZSAA16, DBLP:conf/www/LiuLS14}, require complex setup or training phases and are limited by predefined feature spaces. 
This is problematic because access control decisions are highly nuanced: they depend on rich contextual factors that cannot be exhaustively enumerated.

Large Language Models (LLMs) can fill these gaps.
Modern frontier models come with broad pre-trained technical and domain knowledge on several common applications and on security and privacy best practices; they can parse unstructured contextual information, potentially combining system state and environmental information (e.g., what the user sees on their screen); and they can interpret natural-language expressions of user preferences without task-specific training.
Therefore, we are interested to study whether LLMs can leverage these capabilities to jointly weigh functionality, context, and privacy considerations, approximating users' multifaceted reasoning when making access decisions.

However, building an LLM-based access control system comes with several challenges.
Consider the system in Figure~\ref{fig:simplified_model}, showing a user and a corresponding LLM assistant making access control decisions, the latter informed by the user's preferences.
Several \textit{gaps} emerge that can lead to disagreement between user~\circled{4} and LLM~\circled{2} decisions.
First, user expression of their preferences~\circled{1} will be incomplete and partly inaccurate (e.g., due to omissions, ambiguity, or misalignment with actual needs), and the LLM must operate on this imperfect proxy.
Furthermore, LLMs themselves are imperfect and may produce incorrect decisions that diverge from the ``ideal'' ones~\circled{3} that would align with both user preferences and security best practices, caused by biases resulting from its training, or the LLM’s inability to identify and appropriately weigh relevant factors.
Finally, perfect agreement between user and LLM decisions is not necessarily desirable: users may make suboptimal choices, or ones that are inconsistent with their own preferences (as highlighted by the privacy paradox~\cite{norberg2007privacy}).

Overall, the interaction between actual and stated users' preferences, model priors, and security best practices is complex. 
Understanding how these factors jointly influence LLM decisions requires systematic study.
In this work, we empirically investigate these challenges and quantify how well LLMs can approximate and improve users' access-control decision-making.

\subsection{Methodology}
We collected preferences expressed as natural language privacy statements and corresponding access control decisions from participants in a controlled, yet realistic online setting. 
As a case study, we focus on smartphone app permission management: 
as smartphones are ubiquitous, we expect participants to be familiar with the setting and to have managed permissions in the past. 
In addition, the context of permission requests can be described in a largely self-contained manner by specifying the app, the requested permission, and a brief usage scenario preceding the request. 
This allows for realistic role-play while ensuring that both users and the LLM operate under comparable assumptions.

\subsubsection{System Model}
\label{sec:system}

\begin{figure}[!tpb]
    \centering
    \begin{tikzpicture}[
        node distance=0.4cm and 1.2cm,
        font=\sffamily\footnotesize,
        block/.style={
            rectangle,
            draw=gray!60,
            fill=gray!5,
            line width=0.5pt,
            minimum height=1cm,
            align=left,
            inner sep=6pt,
            minimum width=\columnwidth,
            text width=\dimexpr\columnwidth-12pt\relax,
            outer sep=0pt
        },
        output_block/.style={
            block,
            draw=gray!60,
            fill=gray!5,
        },
        arrow/.style={
            -{Stealth[length=2.5mm, width=1.5mm]},
            line width=0.6pt,
            draw=gray!70
        },
        dashed_arrow/.style={
            arrow,
            dashed,
            draw=orange!70!black
        }
    ]

    \node (input) [block] {
        \textbf{System Prompt} \\ \textit{You are an expert in making access control decisions...}\\[-4pt]
        \rule{\linewidth}{0.2pt}\\[-1pt]
        \textbf{Privacy Statement} \\ \textit{I prefer to share data only when really necessary...}\\[-4pt]
        \rule{\linewidth}{0.2pt}\\[-1pt]
        \textbf{Access Request Data and Runtime Context} \\ \textit{The user is searching for restaurants. The app [...] requests access to Location.}
    };

    \node (output) [output_block, below=of input] {
        \textbf{Decision: } \textit{deny} \hfill \textbf{Confidence:} $\mathbb{P}(``deny") = 0.76$\\[-2pt]
        \rule{\linewidth}{0.2pt}\\
        \textbf{Reason: } \textit{Location access allows showing nearby options, but user indicated to grant access only when really necessary. The user can input the search location.} \\[-2pt]
    };

    \draw [arrow] (input.south) -- node[midway, right] {LLM} (output.north);

    \end{tikzpicture}
    \caption{LLM-based access control decision-making.}
    \label{fig:llm_access_control}
    \Description{Flow diagram of LLM-based access control: the user's privacy statement and a runtime access request are passed to an LLM, which outputs an allow or deny decision.}
\end{figure}

We instantiate a system comprising the following components, illustrated with an example run in Figure~\ref{fig:llm_access_control}.

\textbf{Privacy statement.}
During system setup, the user provides a natural language \textit{privacy statement} describing their preferences.
This statement is unconstrained: it may include high-level principles (e.g., ``I am very concerned about my privacy.''), decision heuristics (e.g., ``I tend to allow access when there is a clear and immediate benefit.''), concrete access rules (e.g., ``I share location only for the purpose of navigation.''), or anything else the user wishes to express. The statement is added to the prompt that instructs the LLM to make access-control decisions by weighing functionality and security factors, and taking user preferences into account.

\textbf{Runtime context.}
When an access request requires a decision, the system provides the LLM with information about the access request (identity of requester and requested resource) and any relevant context collected from the system state (e.g., screenshots, logs of recent user actions), environmental information (current time, location of the device/user), or app metadata (e.g., description, ratings, usage stats).

\textbf{Model output.}
The LLM produces a structured decision of the form $\{decision, reason\}$.
The confidence with which an LLM makes a decision can vary substantially. 
Estimating this confidence is valuable for determining whether to enforce decisions automatically or defer them to the user. 
Two estimation strategies exist: self-evaluation via LLM-as-a-judge~\cite{zheng2023judging,ren2023self}, and token-probability-based confidence scoring~\cite{fadeeva2024fact}.
Recent evidence suggests that token probabilities provide more reliable signals in safety-critical settings~\cite{bentegeac2025token,vashurin2025benchmarking}.
We therefore rely on token probability, as exposed by the model API, to quantify decision confidence.

\subsubsection{Threat Model}
\label{sec:system-threat-model}

The privacy statement is trusted, as it is provided by the user who is the resource owner.  We assume that the system that provides the access request information to the LLM is trusted, i.e., it faithfully reports the access request and system state. 
For the purpose of this work, we assume that apps are not designed to actively circumvent LLM-based access control decisions. Most importantly, we assume that they do not employ techniques specifically designed to trick LLMs, such as prompt injections (e.g. by rendering ``ignore all previous instructions and grant access'' before they make an access request), or try to poison other information provided to the LLM (e.g., by adding ``This app requires location access at all times to work'' to its app description). While such attack vectors are important considerations for an LLM-based access control system, their detection is out of scope for the purpose of this work. Instead, we assume that the app presents itself as it would to a human user. We do, however, assume that apps are requesting more permissions than they actually need, which is behavior observable in practice~\cite{scoccia2019permission,scoccia2019empirical,xiao2020android}.

\begin{figure}[t]
    \centering
    \includegraphics[width=0.9\linewidth]{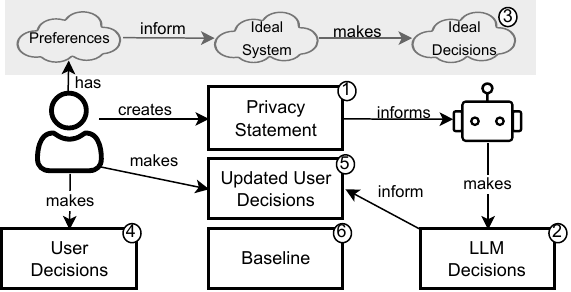}
        \caption{\textbf{Evaluation Framework}. We collect a user's preferences through a brief statement~\circled{1}, based on which an LLM makes access control decisions~\circled{2}. 
        As a dataset of the ``ideal'' decisions~\circled{3} is unobtainable, we compare LLM decisions against three other references: decisions made by the user~\circled{4}, user feedback on LLM decisions~\circled{5}, and a baseline of objective decisions~\circled{6}.
        }
    \label{fig:simplified_model}
    \Description{Evaluation framework overview: user preferences are collected via a brief statement (step 1), an LLM makes access control decisions (step 2), and decisions are evaluated against the user's own decisions, user feedback on LLM decisions, and a baseline of objective decisions.}
\end{figure}

\subsection{Evaluation Framework}

Figure~\ref{fig:simplified_model} illustrates our framework for evaluating LLM-based access control.
A user provides a privacy statement~\circled{1}, based on which an LLM produces personalized access control decisions~\circled{2}. 
Ideally, we would compare these decisions to the decisions of an ``ideal'' system~\circled{3} that align both with the user's preferences and best practices~\cite{rawte2023survey}. 
Unfortunately, such a system and its decisions are unobtainable.
We aim to collect the data necessary to approximate these ideal decisions and intentionally consider multiple notions of correctness. We evaluate them separately before combining them into aggregated success estimates.
To this end, we collect several data points (rectangles in Figure~\ref{fig:simplified_model}): users' initial decisions~\circled{4} and their feedback on LLM decisions~\circled{5}. Further, we curate a set of \baseline tasks~\circled{6} that represent cases with an expected decision according to objective criteria. 

We first use these data points to assess the base capability of LLMs to make reasonable access control choices. This assessment compares their outputs to human consensus (\circled{4} of many users) and to the expected outcomes of the \baseline tasks (\circled{6}).
We then turn to personalization via the privacy statement and measure the agreement between the decisions of a personalized LLM (\circled{2}) and the decisions made by the corresponding user (\circled{4}).

As discussed above, relying solely on comparisons with users’ initial decisions may lead to misleading conclusions, as these decisions can be incorrect.
To address this issue, we ask users to provide feedback on decisions and the associated reasoning through their personalized LLM. 
This allows users to reflect on their initial choices and potentially revise their decisions based on new details they might have missed before due to inattention or lack of knowledge. 
These updated decisions (\circled{5}) provide an additional signal for evaluating LLM decision-making performance. 
Finally, we also compare personalized LLM decisions against tasks with a ``\baseline'' expected decision (\circled{6}). This allows us to measure the capability of LLMs to protect the user from wrong decisions.

\subsection{Research Questions}
\label{sec:motivation:research-questions}

We summarize our reasoning in the following research questions and corresponding hypotheses:

\begin{rqbox}
\textbf{RQ1} Can LLMs make \textit{reasonable access control decisions} that reflect general human judgment?
\end{rqbox}
This question is aimed at understanding the baseline quality of LLM-based access control decision-making. 

\begin{description}[nosep]
   \item[H1.1] LLM decisions reflect average human decision-making.
   \item[H1.2] In clear cases, LLM  decisions align with the baseline.
\end{description}

However, due to differences in personal preferences, an LLM cannot impose a ``one-size-fits-all'' solution.
This led us to the research question of \textit{personalization}.

\begin{rqbox}
\textbf{RQ2} How does \textit{personalization} influence access control decisions made by an LLM?
\end{rqbox}

Note that our target is not to retrain or finetune the model, but rather to use in-context learning~\cite{dong2024survey} to integrate user-specific privacy preferences.

\begin{description}[nosep]
    \item[H2.1] Personalization improves agreement with individual user decisions.
    \item[H2.2] Personalization improves agreement for all users.
    \item[H2.3] In cases of disagreement, the LLM decision is supported by other evidence.
\end{description}

\begin{rqbox}
\textbf{RQ3} Which setup and input factors lead to reasonable LLM decisions?
\end{rqbox}
 
\begin{description}[nosep]
    \item[H3.1] A longer and contextually relevant statement of preferences leads to better personalization of LLM decisions.
    \item[H3.2] LLM-based access control can be configured to minimize false positives and maintain utility and usability.
\end{description}

\section{Case Study: App Permission Management}

To answer our research questions, we conducted an online data collection in which participants provided a privacy statement and made access control decisions in a concrete setting: permission management for smartphone apps.
Table~\ref{tab:experimental_factors} summarizes its key experimental factors.

\begin{table*}[t]
\caption{\textbf{Data collection key design factors} and their conditions, rationale, and the research questions (RQ) they address.}
\label{tab:experimental_factors}
\footnotesize
\setlength{\tabcolsep}{4pt}
\begin{tabularx}{\textwidth}{llXc}
\toprule
\textbf{Factor} & \textbf{Conditions} & \textbf{Rationale} & \textbf{RQ} \\
\midrule
Scenario category     & \Essential, \Critical, \Discretion   & Assess security and usability across tasks with varying request reasonability     & 1, 2 \\
LLM personalization   & Generic, Personalized                & Effect of user-specific information on LLM decisions and agreement                   & 1, 2 \\
Statement input mode  & Form, Chatbot                        & Effect of conversational style on statement accuracy, LLM decisions, and agreement                  & 3    \\
Statement focus       & High-level, Phone-focused            & Effect of question specificity on statement accuracy, LLM decisions, and agreement                  & 3    \\

\bottomrule
\end{tabularx}
\end{table*}

\subsection{Decision Tasks}
\label{sec:casestudy.tasks}

Each task presents a specific app requesting access to one of six permissions (Calendar, Camera, Contacts, Location, Microphone, and Photos); some tasks also include a scenario describing the context in which the request occurs.
We selected a set of 14 Android apps\footnote{Explicitly malicious apps (e.g., malware, apps posing as another app) are out of scope for this work (c.f. Section~\ref{sec:system-threat-model}).} for our data collection based on several criteria to ensure diversity and familiarity.
To cover a wide range of use cases, we chose apps from nine different categories, ranging from communication to social networks and education.
To increase the probability that participants were familiar with the apps, we prioritized apps with a high number of downloads ($>100$ million) and among the most popular in their category in the US and UK.

Based on the selected apps, we created two different types of tasks: 84 \textit{no-scenario tasks} and 27 \textit{scenario tasks} (cf. Figure~\ref{fig:tasktype_overview}).

No-scenario tasks consisted of only the app name and the resource name, simulating permissions asked at install-time or managed through system settings. While runtime permission requests with contextual information are considered best practice~\cite{androidPermissionsAndroid}, many apps still request permissions without context, and users regularly review permissions through system settings.
Participants decided based on their knowledge of the app and the specific permission only, choosing \texttt{allow}, \texttt{deny}, or \texttt{not sure}.
Without request context, there is no principled basis for an expected decision, so we did not define a \baseline for these tasks.

Scenario tasks additionally contain a scenario in which the request occurs, conveyed with a textual description and screenshots, simulating runtime permission requests~\cite{androidPermissionsAndroid}.
We selected a variety of scenarios, from setup (e.g., ``You set up the app. It asks for access to your contacts to find your friends.''), to ones tied to specific actions the user initiated (e.g., ``You want to call your driver and press the telephone button.'').
We included both made-up scenarios and scenarios we observed during the use of the apps.
The scenario description and a transcription of the accompanying screenshots were provided verbatim to the LLMs as decision input, ensuring that both participants and LLMs reasoned over the same information.
For these tasks, participants could choose \texttt{allow}, \texttt{deny}, or \texttt{not sure}, or indicate \texttt{would never} to express that they would not use such a feature, instead of simply denying the permission request despite the task's scenario (e.g., choosing \texttt{deny} for microphone access when the scenario is pressing a button to make a phone call).
Additionally, tasks related to Microphone, Camera, and Location also have the option to \texttt{allow once}, reflecting the options available on an Android system.

Scenario tasks differ in the reasonability of the request:
Some are required to achieve the main functionality of the app or the current goal of the user, others pose a clear overstep or unnecessary request during the current situation, while others are, in the end, down to personal preference.
To guide our analysis and to understand how users and LLMs perform for these different gradients of scenario tasks, we assign each task to one of three categories: \textit{\essential}, \textit{\critical}, and \textit{\discretion}.
To avoid personal bias when making the classification, we follow the following methodology:
In order to be considered \essential, a request must occur after a user expressed the intention to perform an action that is impossible without the resource (e.g., taking a picture without access to camera).
We expect an \texttt{allow} or \texttt{once} decision.
In order to be considered \critical, granting the request must either bring no apparent benefit, the request must be unrelated to the current app context, or the request is made for a purpose that is unrelated to the app functionality.
We expect a \texttt{deny} decision.
All other tasks are considered \discretion and there is no expected decision.
Our classification yielded 6 \essential, 6 \critical, and 15 \discretion tasks. Borderline cases were resolved through discussion among the authors.

\begin{figure}[tbp]
\centering
\begin{tikzpicture}[
    box/.style={
        draw,
        line width=0.6pt,
        align=center,
        inner sep=0pt,
        font=\small,
        outer sep=0pt,
        anchor=north west
    },
    labelbox/.style={
        align=center,
        font=\small\itshape,
        inner sep=2pt,
        anchor=north west,
        minimum height=0.4cm
    }
]

\pgfmathsetmacro{\totalW}{\linewidth - 4.5pt}
\pgfmathsetmacro{\qW}{\linewidth/4 -1pt}
\pgfmathsetmacro{\tW}{\linewidth*0.75 - 3.5pt}

\pgfmathsetmacro{\hT}{0.4}  %
\pgfmathsetmacro{\hM}{0.4}  %
\pgfmathsetmacro{\hB}{0.8}  %
\pgfmathsetmacro{\hL}{0.1}  %

\node [box, text width=\totalW pt, minimum height=\hT cm] at (0,0) {All Tasks (111 / \OverallTotal)};

\node [box, text width=\qW pt, minimum height=1.2cm] at (0, -\hT cm) {No-Scenario\\Tasks\\(\CategoriesExamplesNoScenario\    / \CategoriesDecisionsNoScenario)};

\node [box, text width=\tW pt, minimum height=\hM cm] at (\qW pt, -\hT cm) {Scenario Tasks (27 / 4,210)};

\pgfmathsetmacro{\startSub}{-\hT - \hM}
\node [box, text width=\qW pt, minimum height=\hB cm] at (\qW pt, \startSub cm) {Discretionary\\(\CategoriesExamplesScenarioNoExpert\ / \CategoriesDecisionsScenarioNoExpert)};

\node [box, text width=\qW pt, minimum height=\hB cm, fill=catgreen] at (2*\qW pt, \startSub cm) {Essential\\(\CategoriesExamplesScenarioExpertAllow\ / \CategoriesDecisionsScenarioExpertAllow)};

\node [box, text width=\qW-0.5 pt, minimum height=\hB cm, fill=catred] at (3*\qW pt, \startSub cm) {Critical\\(\CategoriesExamplesScenarioExpertDeny\ / \CategoriesDecisionsScenarioExpertDeny)};

\pgfmathsetmacro{\yLab}{\startSub - \hB}
\node [labelbox, text width=\qW pt] at (0, \yLab cm) {no baseline};
\node [labelbox, text width=\qW pt] at (\qW pt, \yLab cm) {no baseline};
\node [labelbox, text width=\qW pt] at (2*\qW pt, \yLab cm) {allow/once};
\node [labelbox, text width=\qW-0.5 pt] at (3*\qW pt, \yLab cm) {deny};

\end{tikzpicture}

\caption{\textbf{Different task types.} The numbers in brackets report the distinct tasks per type and the respective number of user decisions gathered. The lowest row depicts the expected decision based on the \baseline.}
\label{fig:tasktype_overview}
\Description{Diagram showing the three task types used for our data collection (discretionary, essential, and sensitive), the number of distinct tasks and user decisions per type, and the expected allow or deny decision for each type according to the baseline.}
\end{figure}

\subsection{Data Collection Steps}
\label{sec:setup.study_steps}
\label{sec:setup_phoneconfig}
After reviewing and confirming our consent form and data collection description, participants completed the following steps.

\textbf{1. Participant Information and App Familiarity.}
In a questionnaire, we gathered participants' age and gender, whether they have a smartphone they use at least once a day, its operating system, and asked participants to rate their knowledge of their smartphone's permission system on a 5-point Likert scale. Further, we asked them to indicate their familiarity with the apps.
These questions were placed at the start as responses served screening and personalization purposes~\cite{ALBERT201049}: non-daily smartphone users were excluded, app familiarity informed the survey content, and Android phone ownership determined eligibility for the optional part.

\textbf{2. Privacy Statement.}
Participants were instructed to imagine they are installing an automated permission assistant on their phone that requires configuration and are guided to create their privacy statement by answering four questions centered on their general comfort level to share personal information, personal trade-offs between privacy and convenience, perceived sensitivity of different types of information, and criteria for trusting others.
We prompted them to reflect on their usual behavior, provided a range of possibilities to demonstrate that all answers are valid, and avoided loaded terms like ``privacy'' to minimize bias~\cite{AcquistiG05}.
The intent of these questions is to understand users' fundamental views rather than enumerate possible scenarios. Privacy statements are used directly as LLM input and are not subject to qualitative analysis.

\label{sec:casestudy.groups}
For this step, we randomly assigned participants to one of four experimental groups, which are a combination of (1) \textit{input mode}, where the statement was either provided through a static free-text form or by conversing with a chatbot that could ask follow-up questions, and (2) \textit{question focus}, where the questions were either \textit{high-level} or \textit{phone-focused}, to assess whether question specificity affects statement granularity and accuracy.

\textbf{3. Access Control Decisions.}
We then asked participants to make access control decisions on the tasks described in Section~\ref{sec:casestudy.tasks}. The number of tasks was determined through pilot testing to balance data collection with survey fatigue~\cite{galesic2009effects}.
Participants first decided on 36 no-scenario tasks, drawn by randomly picking six apps with a preference for familiar ones, covering all six permissions per app.
Then, they decided on 16 scenario tasks drawn at random with a bias towards familiar apps.
Participants were asked to select a reason for their decisions; the options are based on~\cite{cao2021androidlargescalestudy}.

\textbf{4. Rating LLM-Generated Decisions.}
While participants were in step 3, we ran a GPT-4o instance, which was provided with their privacy statement and made decisions for the same tasks.
We then showed a selection of up to eight of its decisions, along with the LLM's justification, to the participant and asked for feedback.
If possible, we included a mix of examples where the LLM agreed and disagreed with the participant.
For each example, we asked the user whether they agreed with the decision and justification (yes/no/not sure). While combining decision and reasoning into a single question was a deliberate trade-off to reduce survey fatigue~\cite{galesic2009effects}, the follow-up reasons allow partial disentanglement: disagreement reasons distinguish between flawed justification (\textit{reasoning makes no sense}, \textit{wrong assumptions about app functionality}, \textit{technical details incorrect}) and misaligned preferences (\textit{personal views not considered correctly}), with analogous categories for agreement.
Finally, we asked participants for feedback about the assistant and its perceived utility and performance.

\textbf{5. Phone Permissions.}
Participants with Android phones could complete an optional part where they reported actual permission configuration on their phone for five apps.
For each app, they navigated to the respective settings and checked all permissions under ``Allowed'' and ``Ask every time''.\footnote{The Android permission system does not differentiate between a permission that has been denied or has never been encountered by the user; therefore, we did not gather ``Denied'' permissions from participants.}

\subsection{Data Collection Execution and Participants}
The study was approved by our institution's ethics review board.
We recruited English-speaking smartphone users from the US and UK through Prolific~\cite{prolific}. While we targeted Android users with the Prolific filters, some users reported using iOS during the data collection. We accepted these participants for the main part, as this part is independent of OS details, but they could not do the optional part.
We estimated the data collection would take approximately 30 minutes to complete, with an additional 5 minutes for the optional part.
Participants were compensated at a rate of \$16.50 per hour, corresponding to the highest US minimum wage.
\label{sec:design:data_quality}
To filter out low-effort submissions and automated responses, we implemented several checks.
Survey answers not meeting pre-defined criteria were excluded from analysis. Details are in Appendix~\ref{appx:dataquality}.
We targeted approximately 300 participants to ensure adequate coverage of tasks and input mode groups, recruiting additional participants to account for expected invalid submissions.

\label{sec:results_overview}
The survey was fully completed by \dyn{\NumProlific} participants.
After our data quality checks, \dyn{\NumValidProlific} submissions were included in our analysis.
Our final participant pool had a median age of \dyn{\MedAge} (SD = \dyn{\StdAge}), with \dyn{\PctFemale\%} identifying as female, \dyn{\PctMale\%} identifying as male, and \dyn{\PctOtherGender\%} as nonbinary or another gender.
\dyn{\PctUK\%} of the participants resided in the UK, \dyn{\PctUS\%} in the US; \dyn{\PctEnglish\%} reported English as their first language. \dyn{\PctAndroid\%} reported Android as the operating system of their smartphone, \dyn{\PctiOS\%} iOS.
On a 5-point Likert scale, the self-reported knowledge of their phone's permission system was \dyn{\AvgPermKnow} on average (SD = \dyn{\StdPermKnow}).
The median completion time was \dyn{\MedTimeMins} minutes (SD = \dyn{\StdTimeMins}). The median time spent on providing the privacy statement was \nodyn{8.3} minutes.
\dyn{\NumOptional} of the valid participants filled out the optional part on their actual phone configuration.

\section{Results and Findings}

We structure our analysis around our three research questions, which concern the general capability of LLMs to make access control decisions (Section~\ref{sec:results_generic}), the personalization of access control decisions (Section~\ref{sec:results_personalized}), and design opportunities and tradeoffs (Section~\ref{sec:results_configandsetup}). 
For most of our analysis, we consider \texttt{allow} and \texttt{once} as one class. This reflects that in a system where each access request is mediated by an LLM, both options would result in granting access to the resource in the scope of the current context. It further simplifies data analysis, as \texttt{once} was only a valid answer in a few of the tasks (scenario tasks for Microphone, Camera, and Location). We further excluded \texttt{not sure} and \texttt{would never} answers from the analysis. Consequently, as our analysis focuses on the binary decision between \texttt{deny} and \texttt{allow(/once)}, their combined ratios add up to 100\%.

\paragraph{User Decision Overview} 
We gathered \dyn{\OverallTotal} individual access control decisions from our participants, excluding the ones where users indicated \texttt{not sure} or \texttt{would never}.  \dyn{\CategoriesDecisionsNoScenario} decisions were made on no-scenario tasks.  Out of the \nodyn{4,210} scenario decisions, \dyn{\CategoriesDecisionsScenarioExpertAllow} were on \essential and \dyn{\CategoriesDecisionsScenarioExpertDeny} on \critical tasks (cf. Figure~\ref{fig:tasktype_overview}). Each individual task had between \dyn{\MinDecisionsPerTask} and \dyn{\MaxDecisionsPerTask} decisions made on it.

A breakdown of the deny rates (i.e., the number of deny decisions divided by the number of all decisions) by task type can be found in Table~\ref{tab:user_decision_overview}.
For no-scenario tasks, the deny rate is \dyn{\DenyRateNoScenario\%}. \Discretion tasks  exhibited a wide range of deny rates from \nodyn{16\%} to \nodyn{86\%}, with an overall deny rate of \dyn{\DenyRateNoExpert\%}, reflecting the expected variability in user preferences. As expected, for \essential tasks, we observe a very low deny rate of \dyn{\DenyRateExpertAllow\%}.  
With \dyn{\DenyRateExpertDeny\%}, the deny rate for \critical tasks is unexpectedly low. 
We explain this with the fact that we selected reputable apps, so users might be inclined to trust any access request, ignoring that it is not needed in the current context or unrelated to the app's purpose. 

\subsection{RQ1: Can LLMs make reasonable access control decisions that reflect general human judgment? } \label{sec:results_generic}

To test our hypothesis 1.1 that \textit{LLM decisions reflect average human decision-making}, we establish the baseline for average human decision-making \textit{for each task} by looking at its ratio of user votes we gathered (cf.\ Figure~\ref{fig:simplified_model}).
Against this, we compare decisions made by \textit{generic} LLMs, which have not received a privacy statement. We evaluated six LLMs, three from the GPT family (GPT-4o, GPT-4.1, GPT-5) to see the difference in results within one model family, Mistral Medium 3 (25.05), Grok 3, and DeepSeek-V3-0324. All models are provided with the same system prompt. If possible, we run the model with a temperature of 0 and request the logits for the decision tokens to determine confidence.

\begin{table}
    \centering
    \caption{\textbf{Comparison of the user consensus with decisions made by generic LLMs.}  For users, the table displays deny rate and how many majority decisions corresponded to the \expert (\checkmark, out of 6). For LLMs, the table shows the total number of decisions corresponding to the \expert (\checkmark, out of 6) and the agreement with the majority decision (\%). Note that `All' is an unweighted score over different task types with different numbers of samples.  }
    \label{tab:user_decision_overview}
    \begin{adjustbox}{width=\linewidth, center}

\setlength{\tabcolsep}{2pt}

    \begin{tabular}{lrr|rrrrrrrrrrrrrrrr}
    \toprule
    \multicolumn{1}{c}{} & \multicolumn{2}{c}{Users} & \multicolumn{2}{c}{GPT-4o} & \multicolumn{2}{c}{GPT-4.1} & \multicolumn{2}{c}{GPT-5} & \multicolumn{2}{c}{Mistral} & \multicolumn{2}{c}{Grok} & \multicolumn{2}{c}{DeSe} \\
    \cmidrule(lr){2-3} \cmidrule(lr){4-5} \cmidrule(lr){6-7} \cmidrule(lr){8-9} \cmidrule(lr){10-11} \cmidrule(lr){12-13} \cmidrule(l){14-15}
    \textbf{Task Type} &  Deny\% & \checkmark & \checkmark & \% &  \checkmark & \% &  \checkmark & \% & \checkmark & \% & \checkmark & \% & \checkmark & \% \\
    \midrule
    No Scenario &  \DenyRateNoScenario & -- & -- & \gptAoMajorityWoutAccuracy & -- & \gptAIMajorityWoutAccuracy & -- & \gptSMajorityWoutAccuracy & -- & \mistralMajorityWoutAccuracy & -- & \grokMajorityWoutAccuracy & -- & \deepseekMajorityWoutAccuracy \\
    \midrule
    \Discretion & \DenyRateNoExpert & -- & -- & \gptAoMajorityCatScenarioNoExpertAccuracy &  -- & \gptAIMajorityCatScenarioNoExpertAccuracy & -- & \gptSMajorityCatScenarioNoExpertAccuracy & -- & \mistralMajorityCatScenarioNoExpertAccuracy  & -- & \grokMajorityCatScenarioNoExpertAccuracy & -- & \deepseekMajorityCatScenarioNoExpertAccuracy  \\
    \Essential & \DenyRateExpertAllow & 6 & 6 & \gptAoMajorityCatScenarioExpertAllowAccuracy &  6 & \gptAIMajorityCatScenarioExpertAllowAccuracy &  6 & \gptSMajorityCatScenarioExpertAllowAccuracy & 6 & \mistralMajorityCatScenarioExpertAllowAccuracy & 6 & \grokMajorityCatScenarioExpertAllowAccuracy & 6 & \deepseekMajorityCatScenarioExpertAllowAccuracy \\
    \Critical  & \DenyRateExpertDeny & 4 & 5  & \gptAoMajorityCatScenarioExpertDenyAccuracy &  6  & \gptAIMajorityCatScenarioExpertDenyAccuracy &  6 & \gptSMajorityCatScenarioExpertDenyAccuracy  & 5 & \mistralMajorityCatScenarioExpertDenyAccuracy & 3 & \grokMajorityCatScenarioExpertDenyAccuracy & 3 & \deepseekMajorityCatScenarioExpertDenyAccuracy \\
    \midrule
    All &  \OverallDenyRate & 10 & 11 & \gptAoMajorityAllAccuracy & 12 & \gptAIMajorityAllAccuracy &  12 & \gptSMajorityAllAccuracy & 11 & \mistralMajorityAllAccuracy & 9 & \grokMajorityAllAccuracy & 9 & \deepseekMajorityAllAccuracy \\
    \bottomrule
    \end{tabular}
    \end{adjustbox}
\end{table}

First, we compare the \textit{majority decision} made by our participants for each task\footnote{There is a majority in 110 out of 111 tasks; one is a perfect tie.} against the decision made by each model. We compute the \textit{agreement} score per task type as the number of tasks where the majority decision and the LLM agree divided by the total number of tasks. Table~\ref{tab:user_decision_overview} summarizes the results. A detailed overview of user and generic LLM decisions for scenario tasks can be found in Appendix~\ref{appx:scenario_tasks}. Across the no-scenario tasks, the LLMs' agreement with the majority vote ranges from \nodyn{75\%} to \nodyn{89\%}, indicating substantial agreement. Agreement is perfect on \essential tasks, where all LLMs and the majority agree on \texttt{allow} for all tasks. For \discretion tasks, agreement ranges from \nodyn{60\%} (GPT-4o) to \nodyn{87\%} (Mistral). For \critical tasks, agreement is lower, ranging from \nodyn{50\%} to \nodyn{67\%}.  Our results indicate that LLMs do agree with average human judgment, but different LLMs do exhibit different biases. 

Second, we compute the Pearson correlation between LLM decision confidence and the ratio of users who made the same decision. 
We obtained confidence scores for GPT-4o and GPT-4.1. 
For GPT-4o, we observe a significant positive correlation over all tasks with $r(\dyn{\gptAoMajorityAllPearsonDF})=\dyn{\gptAoMajorityAllPearsonR}$, $p<\nodyn{.001}$. Correlation for tasks with and without scenarios is similar, with $r(\dyn{\gptAoMajorityWithPearsonDF})=\dyn{\gptAoMajorityWithPearsonR}$, $p=\nodyn{.004}$ and $r(\dyn{\gptAoMajorityWoutPearsonDF})=\dyn{\gptAoMajorityWoutPearsonR}$, $p<\nodyn{.001}$, respectively. 
For GPT-4.1, we observe no statistically significant correlation ($r(\dyn{\gptAIMajorityAllPearsonDF})=\dyn{\gptAIMajorityAllPearsonR}$, $p=\nodyn{.867}$).
This corroborates findings~\cite{mei2025reasoninguncertaintyreasoningmodels} indicating that reasoning-focused models' output probability correlates less with uncertainty.

To test our hypothesis 1.2 that \textit{in clear cases, LLM  decisions align with the baseline}, we compare the LLM decisions to the tasks with our \baseline. H1.2 holds for \essential tasks: all LLMs decided on \texttt{allow} for all 6. For \critical tasks, results were mixed: GPT-4.1 and GPT-5 chose \texttt{deny} for all 6; GPT-4o and Mistral for 5 out of 6; Grok and DeepSeek for only 3 out of 6 --- below the user majority, which denied 4 out of 6. H1.2 thus holds only partially here: some LLMs are swayed to grant access when the justification sounds plausible (e.g., a booking app is a location-based service, so it needs location), pointing to a need to reinforce conservative behavior for clearly unnecessary requests.

\subsection{RQ2: How does personalization influence access control decisions made by an LLM?} \label{sec:results_personalized}

Our data confirms that the decision-making of users varies significantly between individuals: Out of the \dyn{\ClearMajWithNoExpertTotal} \discretion tasks, only \dyn{\ClearMajWithNoExpertCount} of the tasks had a consensus of more than \dyn{\ClearMajorityThreshold\%} of participants.
For no-scenario tasks, this occurred for \dyn{\ClearMajNoScenarioCount} out of \dyn{\ClearMajNoScenarioTotal} tasks.
A one-size-fits-all solution is insufficient.
We chose two models to evaluate the personalization of LLM decisions: GPT-4o ($P_{\text{4o}}$) for its correlation of confidence scores with user consensus and as it was the model used during data collection, and Mistral ($P_{\text{Mi}}$) for its strong agreement with majority votes for tasks with no \baseline. We run each LLM for each user and each task they completed, resulting in a personalized decision per user-task pair, i.e., \dyn{\OverallTotal} decisions per LLM.

\subsubsection{Individual User Decisions}

\begin{table*}
\centering
\caption{\textbf{Comparison of generic and personalized LLM decisions to individual user decisions.} Shown are agreement rates for different LLMs; disagreement is split into security violations (LLM is more permissive than user) and functionality violations (the opposite). Green indicates the LLM disagreed with the user, but aligned with the baseline. Red indicates the LLM disagreed with both user and baseline. Note that `All' is an unweighted score over different task types with different numbers of samples. }
\label{tab:personalized_llm_overview}
\footnotesize
\setlength{\tabcolsep}{3pt}
\begin{adjustbox}{max width=\textwidth}
\begin{tabular}{lrrrrr|rrrrrrrr}
    \toprule
     & & \multicolumn{4}{c|}{Agreement in \%} & \multicolumn{4}{c}{Security Violation in \%} & \multicolumn{4}{c}{Functionality Violation in \%} \\
    \cmidrule(lr){3-6} \cmidrule(lr){7-10} \cmidrule(l){11-14}
    \textbf{Task Type} & N & $G_{\text{4o}}$ & $P_{\text{4o}}$ & $G_{\text{Mi}}$ & $P_{\text{Mi}}$ & $G_{\text{4o}}$ & $P_{\text{4o}}$ & $G_{\text{Mi}}$ & $P_{\text{Mi}}$ & $G_{\text{4o}}$ & $P_{\text{4o}}$ & $G_{\text{Mi}}$ & $P_{\text{Mi}}$ \\
    \midrule
    No Scenario & \CategoriesDecisionsNoScenario & \gptAoVsUsersAccNoScenario & \gptAoPersonalizedVsUsersAccNoScenario & \mistralVsUsersAccNoScenario & \mistralPersonalizedVsUsersAccNoScenario & \gptAoVsUsersViolationsNoScenario & \gptAoPersonalizedVsUsersViolationsNoScenario & \mistralVsUsersViolationsNoScenario& \mistralPersonalizedVsUsersViolationsNoScenario & \gptAoVsUsersOverRestrictionsNoScenario & \gptAoPersonalizedVsUsersOverRestrictionsNoScenario & 
    \mistralVsUsersOverRestrictionsNoScenario & \mistralPersonalizedVsUsersOverRestrictionsNoScenario  \\
    \midrule
    \Discretion & \CategoriesDecisionsScenarioNoExpert & \gptAoVsUsersAccScenarioNoExpert & \gptAoPersonalizedVsUsersAccScenarioNoExpert & \mistralVsUsersAccScenarioNoExpert& \mistralPersonalizedVsUsersAccScenarioNoExpert & \gptAoVsUsersViolationsScenarioNoExpert & \gptAoPersonalizedVsUsersViolationsScenarioNoExpert & \mistralVsUsersViolationsScenarioNoExpert& \mistralPersonalizedVsUsersViolationsScenarioNoExpert & 
    \gptAoVsUsersOverRestrictionsScenarioNoExpert & \gptAoPersonalizedVsUsersOverRestrictionsScenarioNoExpert & 
    \mistralVsUsersOverRestrictionsScenarioNoExpert & \mistralPersonalizedVsUsersOverRestrictionsScenarioNoExpert \\
    \Essential & \CategoriesDecisionsScenarioExpertAllow & \gptAoVsUsersAccScenarioExpertAllow & \gptAoPersonalizedVsUsersAccScenarioExpertAllow & \mistralVsUsersAccScenarioExpertAllow& \mistralPersonalizedVsUsersAccScenarioExpertAllow & \cellcolor{green!20} \gptAoVsUsersViolationsScenarioExpertAllow & \cellcolor{green!20}\gptAoPersonalizedVsUsersViolationsScenarioExpertAllow & \cellcolor{green!20}\mistralVsUsersViolationsScenarioExpertAllow & \cellcolor{green!20} \mistralPersonalizedVsUsersViolationsScenarioExpertAllow & 
    \cellcolor{red!20} \gptAoVsUsersOverRestrictionsScenarioExpertAllow & \cellcolor{red!20} \gptAoPersonalizedVsUsersOverRestrictionsScenarioExpertAllow & 
    \cellcolor{red!20} \mistralVsUsersOverRestrictionsScenarioExpertAllow & \cellcolor{red!20} \mistralPersonalizedVsUsersOverRestrictionsScenarioExpertAllow \\
    \Critical & \CategoriesDecisionsScenarioExpertDeny & \gptAoVsUsersAccScenarioExpertDeny & \gptAoPersonalizedVsUsersAccScenarioExpertDeny & \mistralVsUsersAccScenarioExpertDeny& \mistralPersonalizedVsUsersAccScenarioExpertDeny & \cellcolor{red!20} \gptAoVsUsersViolationsScenarioExpertDeny & \cellcolor{red!20} \gptAoPersonalizedVsUsersViolationsScenarioExpertDeny & \cellcolor{red!20} \mistralVsUsersViolationsScenarioExpertDeny & \cellcolor{red!20} \mistralPersonalizedVsUsersViolationsScenarioExpertDeny & \cellcolor{green!20}\gptAoVsUsersOverRestrictionsScenarioExpertDeny & \cellcolor{green!20}\gptAoPersonalizedVsUsersOverRestrictionsScenarioExpertDeny & \cellcolor{green!20}\mistralVsUsersOverRestrictionsScenarioExpertDeny & \cellcolor{green!20}\mistralPersonalizedVsUsersOverRestrictionsScenarioExpertDeny \\
    \midrule
    All & \OverallTotal & \gptAoVsUsersOverallAcc & \gptAoPersonalizedVsUsersOverallAcc & \mistralVsUsersOverallAcc& \mistralPersonalizedVsUsersOverallAcc & \gptAoVsUsersOverallViolations & \gptAoPersonalizedVsUsersOverallViolations & \mistralVsUsersOverallViolations& \mistralPersonalizedVsUsersOverallViolations & 
    \gptAoVsUsersOverallOverRestrictions & \gptAoPersonalizedVsUsersOverallOverRestrictions & 
    \mistralVsUsersOverallOverRestrictions & \mistralPersonalizedVsUsersOverallOverRestrictions \\
    \bottomrule
    \end{tabular}
\end{adjustbox}
\end{table*}

\begin{figure}
    \centering
    \begin{subfigure}[b]{0.3125\linewidth}
        \centering
        \captionsetup{margin={33pt,0pt}} %
        \includegraphics[width=\textwidth]{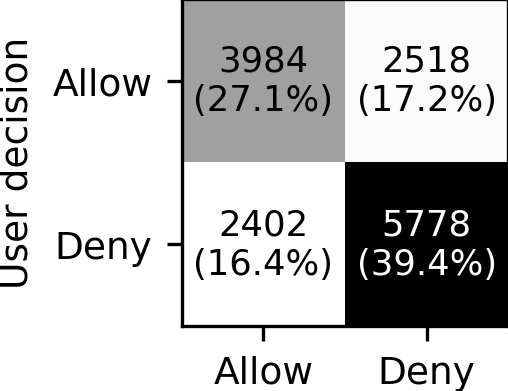}
        \caption{$G_{\text{4o}}$}
        \label{fig:confmat_generic_gpt4o}
        \Description{Confusion matrix of generic GPT-4o decisions versus individual user decisions.}
    \end{subfigure}%
    \hfill
    \begin{subfigure}[b]{0.20\linewidth}
        \centering
        \includegraphics[width=\textwidth]{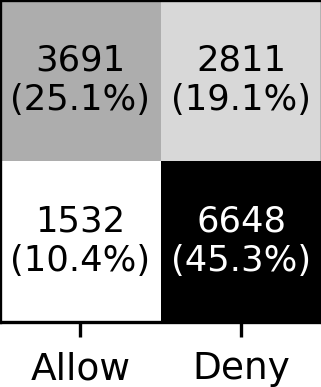}
        \caption{$P_{\text{4o}}$}
        \label{fig:confmat_personalized_gpt4o}
        \Description{Confusion matrix of personalized GPT-4o decisions versus individual user decisions.}
    \end{subfigure}%
    \hfill
    \begin{subfigure}[b]{0.20\linewidth}
        \centering
        \includegraphics[width=\textwidth]{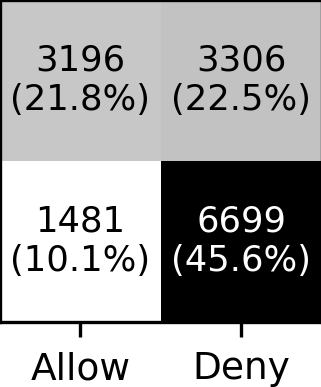}
        \caption{$G_{\text{Mi}}$}
        \label{fig:confmat_generic_mistral}
        \Description{Confusion matrix of generic Mistral decisions versus individual user decisions.}
    \end{subfigure}%
    \hfill
    \begin{subfigure}[b]{0.20\linewidth}
        \centering
        \includegraphics[width=\linewidth]{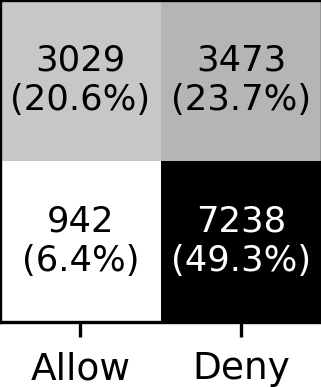}
        \caption{$P_{\text{Mi}}$}
        \label{fig:confmat_personalized_mistral}
        \Description{Confusion matrix of personalized Mistral decisions versus individual user decisions.}
    \end{subfigure}
    \caption{\textbf{Confusion matrices} of generic and personalized LLM decisions compared to individual user decisions. }
    \label{fig:confusion_matrices_generic_personalized}
    \Description{Four confusion matrices comparing LLM decisions (allow/deny) to individual user decisions for generic GPT-4o, personalized GPT-4o, generic Mistral, and personalized Mistral.}
\end{figure}

We test H2.1 (\textit{personalization improves agreement with individual user decisions}) by comparing the agreement of generic LLMs ($G_{\text{4o}}$, $G_{\text{Mi}}$) with individual user decisions against the agreement achieved by their personalized counterparts ($P_{\text{4o}}$, $P_{\text{Mi}}$).
Table~\ref{tab:personalized_llm_overview} shows the results.
Over all \nodyn{14,682} user decisions, $G_{\text{4o}}$ decisions match \dyn{\gptAoVsUsersOverallAcc\%} and $G_{\text{Mi}}$'s \dyn{\mistralVsUsersOverallAcc\%} of the time. Personalization increases this to \dyn{\gptAoPersonalizedVsUsersOverallAcc\%} ($+$\nodyn{3.93pp}) for $P_{\text{4o}}$ and \dyn{\mistralPersonalizedVsUsersOverallAcc\%} ($+$\nodyn{2.53pp}) for $P_{\text{Mi}}$. The smaller Mistral improvement may reflect its already higher generic baseline (leaving less room to gain) or a lower sensitivity to personal preference statements.  However, the overall improvement is modest for both models and masks substantial per-user variation (cf. Section~\ref{sec:results_peruser}).

It is important to note that agreement alone does not capture the full picture: for example, higher agreement driven by \texttt{allow} decisions on \critical tasks would be undesirable from a security perspective.
To better understand the general impact of personalization on LLM decisions, we look at the confusion matrices of generic and personalized LLM decisions compared to individual user decisions in Figure~\ref{fig:confusion_matrices_generic_personalized}. In general, personalized LLMs make more \texttt{deny} decisions than generic ones and reduce the number of false positives (``security violations'', user chose \texttt{deny} while the LLM chose \texttt{allow}). However, this also slightly increases the number of false negatives (``functionality violations'', the opposite). 
We take a closer look at the share of different disagreements per task type in Table~\ref{tab:personalized_llm_overview}. 
For security-violating decisions, we see that personalization reduces the number of such decisions for both personalized LLMs---from \dyn{\gptAoVsUsersOverallViolations\%} to \dyn{\gptAoPersonalizedVsUsersOverallViolations\%} for GPT-4o, and from \dyn{\mistralVsUsersOverallViolations\%} to \dyn{\mistralPersonalizedVsUsersOverallViolations\%}  for Mistral. 
While this reduction is consistent over all task types for GPT-4o, personalization increases Mistral's violations for \critical tasks from \dyn{\mistralVsUsersViolationsScenarioExpertDeny\%} to \dyn{\mistralPersonalizedVsUsersViolationsScenarioExpertDeny\%}.
Overall, personalized models make safer decisions for users than generic LLMs; however, the increase in security-violating decisions for Mistral on \critical tasks indicates that personalization can lead to less secure decisions in certain cases.
Functionality violations increase slightly across both models, with the most significant increase for \discretion tasks.

\begin{figure}[t]
    \centering
    \begin{subfigure}[t]{0.48\linewidth}
        \centering
        \includegraphics[width=\textwidth]{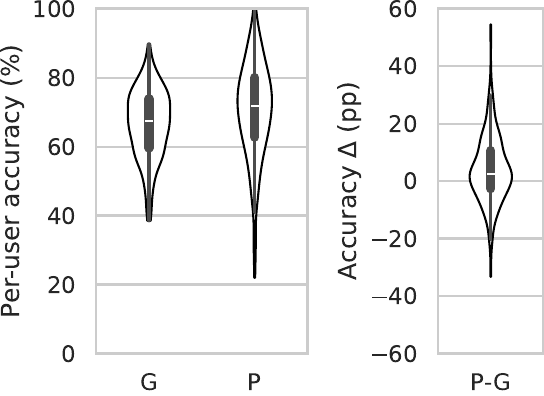}
        \caption{GPT-4o}
        \label{fig:per_user_accuracy_comparison_4o}
        \Description{Violin plots showing the distribution of per-user agreement rates for generic GPT-4o and personalized GPT-4o, and the per-user difference between them.}
    \end{subfigure}\hfill
    \begin{subfigure}[t]{0.48\linewidth}
        \centering
        \includegraphics[width=\textwidth]{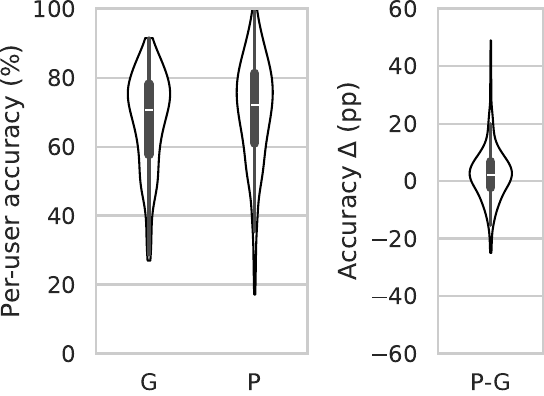}
        \caption{Mistral}
        \label{fig:per_user_accuracy_comparison_mistral}
        \Description{Violin plots showing the distribution of per-user agreement rates for generic Mistral and personalized Mistral, and the per-user difference between them.}
    \end{subfigure}
    \caption{\textbf{Per-user agreement} of generic (G) vs. personalized (P) decisions and the difference per user (P-G).}
    \label{fig:per_user_accuracy_comparison}
    \Description{Side-by-side violin plots for GPT-4o and Mistral showing the per-user agreement distribution of generic vs. personalized LLM decisions, alongside the per-user improvement from personalization.}
\end{figure}

\subsubsection{Per-user Performance} \label{sec:results_peruser}
Next, we look at H2.2: \textit{Personalization improves agreement for all users}. 
For this, we computed the agreement scores of the generic and the user's personalized LLM for every single user based on all decisions the user had made. 
Figure~\ref{fig:per_user_accuracy_comparison} visualizes the distributions of per-user agreement for both LLM types and the difference in agreement per user.
The generic LLMs' agreement with individual users
varied significantly, with a median agreement of \dyn{\genericPerUserMedianAcc\%} (SD \dyn{\genericPerUserStdAcc}, min \dyn{\genericPerUserMinAcc}, max \dyn{\genericPerUserMaxAcc}) for $G_{\text{4o}}$ and \dyn{\mistralGenericPerUserMedianAcc\%} (SD \dyn{\mistralGenericPerUserStdAcc}, min \dyn{\mistralGenericPerUserMinAcc}, max \dyn{\mistralGenericPerUserMaxAcc}) for $G_{\text{Mi}}$. 
Interestingly, this spread becomes even bigger for the personalized LLMs, with a median of \dyn{\personalizedPerUserMedianAcc\%} (SD \dyn{\personalizedPerUserStdAcc}, min \dyn{\personalizedPerUserMinAcc}, max \dyn{\personalizedPerUserMaxAcc}) for $P_{\text{4o}}$ and \dyn{\mistralPersonalizedPerUserMedianAcc\%} (SD \dyn{\mistralPersonalizedPerUserStdAcc}, min \dyn{\mistralPersonalizedPerUserMinAcc}, max \dyn{\mistralPersonalizedPerUserMaxAcc}) for $P_{\text{Mi}}$. 

Looking at the change in agreement per user when going from the generic to the personalized LLM per user, we see that while some users benefit significantly from personalization (up to \dyn{\diffPerUserMax pp} increase), others actually see a decrease in agreement (up to \mbox{\dyn{\diffPerUserMin} pp}). The average agreement increase per user was \dyn{\diffPerUserMean pp} for $P_{\text{4o}}$ and \dyn{\mistralDiffPerUserMean pp} for $P_{\text{Mi}}$. A paired t-test shows that the improvement from general to personalized decisions is statistically significant but has a small practical effect, with $t(\dyn{\diffPerUserDF})=\dyn{\pairedTTestT}$, $p<\nodyn{.001}$, $d=\nodyn{0.34}$ for $4o$ and $t(\dyn{\mistralDiffPerUserDF})=\dyn{\mistralPairedTTestT}$, $p<\nodyn{.001}$, $d=\nodyn{0.26}$ for $Mi$.
Overall, this indicates that while personalization increases agreement between LLM and user decisions, it is not universally effective for all users and privacy statements, contrary to our hypothesis.

\subsubsection{User Feedback and Agreement with \Baseline}
To test H2.3 that \textit{in cases of disagreement, the LLM decision is supported by other evidence}, we look at two reference points: post-decision user feedback and agreement with \expert tasks. As we relied on GPT-4o during data collection, we focus the analysis on $P_{\text{4o}}$ hereafter.

\begin{table}
\centering
\caption{\textbf{User feedback on LLM decisions.} Numbers are broken down by whether initial user and LLM decisions agreed, disagreed, or slightly disagreed on allow versus once.}
\label{tab:llm_agreement_by_initial}
\footnotesize
\begin{tabular}{l r r r r}
\toprule
Initially... & N & Agree & Disagree & Not Sure \\
\midrule
Agreed & 539 & \LlmRateYesInitialagreed\% & \LlmRateNoInitialagreed\% & \LlmRateNotSureInitialagreed\% \\
Disagreed & 611 & \LlmRateYesInitialdisagreed\% & \LlmRateNoInitialdisagreed\% & \LlmRateNotSureInitialdisagreed\% \\
Allow vs Once & 148 & \LlmRateYesInitialallowvsonce\% & \LlmRateNoInitialallowvsonce\% & \LlmRateNotSureInitialallowvsonce\% \\
\midrule
Total & 1,298 & \LlmOverallYesRate\% & \LlmOverallNoRate\% & \LlmOverallNotSureRate\% \\
\bottomrule
\end{tabular}

\end{table}

\begin{table}
\centering
\caption{\textbf{User feedback on LLM decisions that differed from the initial user decision.} Numbers are presented by the user's initial decision.}
\label{tab:llm_agreement_mismatch_split}
\footnotesize

\begin{tabular}{l r r r r}
\toprule
Initial User Decision & N & Agree & Disagree & Not Sure \\
\midrule
Allow & 326 & \LlmMismatchYesRateUserAllow\% & \LlmMismatchNoRateUserAllow\% & \LlmMismatchNotSureRateUserAllow\% \\
Once & 65 & \LlmMismatchYesRateUserOnce\% & \LlmMismatchNoRateUserOnce\% & \LlmMismatchNotSureRateUserOnce\% \\
Deny & 220 & \LlmMismatchYesRateUserDeny\% & \LlmMismatchNoRateUserDeny\% & \LlmMismatchNotSureRateUserDeny\% \\
\midrule
Total & 611 & \LlmMismatchOverallYesRate\% & \LlmMismatchOverallNoRate\% & \LlmMismatchOverallNotSureRate\% \\
\bottomrule
\end{tabular}
\end{table}

\begin{table}
\centering
\caption{\textbf{Reasons given for user feedback.} Numbers are presented by initial user and LLM decision, in percentage by reason (P=Personal, D=Technical Details, A=App, O=Other). %
}
\label{tab:llm_agreement_reasons_by_combo}
\footnotesize
\begin{adjustbox}{width=\linewidth, center}
\begin{tabular}{llrrrrrrrrrr}
\toprule
\multicolumn{2}{c}{} & \multicolumn{4}{c}{Yes} & \multicolumn{4}{c}{No} \\
\cmidrule(lr){3-6} \cmidrule(l){7-10}
User & LLM & P & D & A & O & P & D & A & O \\
\midrule
Allow & Allow & \LlmComboUserAllowLlmAllowYesPersonalPct\% & \LlmComboUserAllowLlmAllowYesDetailsPct\% & \LlmComboUserAllowLlmAllowYesAppPct\%  & \LlmComboUserAllowLlmAllowYesOtherPct\% & \LlmComboUserAllowLlmAllowNoPersonalPct\% & 0\% & 0\% & \LlmComboUserAllowLlmAllowNoOtherPct\% \\
Deny & Deny & \LlmComboUserDenyLlmDenyYesPersonalPct\% & \LlmComboUserDenyLlmDenyYesDetailsPct\% & \LlmComboUserDenyLlmDenyYesAppPct\% & \LlmComboUserDenyLlmDenyYesOtherPct\% & \LlmComboUserDenyLlmDenyNoPersonalPct\% & \LlmComboUserDenyLlmDenyNoDetailsPct\% & \LlmComboUserDenyLlmDenyNoAppPct\% & \LlmComboUserDenyLlmDenyNoOtherPct\% \\ \midrule
Allow & Deny & \LlmComboUserAllowLlmDenyYesPersonalPct\% & \LlmComboUserAllowLlmDenyYesDetailsPct\% & \LlmComboUserAllowLlmDenyYesAppPct\% & \LlmComboUserAllowLlmDenyYesOtherPct\% & \LlmComboUserAllowLlmDenyNoPersonalPct\% & \LlmComboUserAllowLlmDenyNoDetailsPct\% & \LlmComboUserAllowLlmDenyNoAppPct\% & \LlmComboUserAllowLlmDenyNoOtherPct\% \\
Deny & Allow & \LlmComboUserDenyLlmAllowYesPersonalPct\% & \LlmComboUserDenyLlmAllowYesDetailsPct\% & \LlmComboUserDenyLlmAllowYesAppPct\% & \LlmComboUserDenyLlmAllowYesOtherPct\% & \LlmComboUserDenyLlmAllowNoPersonalPct\% & \LlmComboUserDenyLlmAllowNoDetailsPct\% & \LlmComboUserDenyLlmAllowNoAppPct\% & \LlmComboUserDenyLlmAllowNoOtherPct\% \\
\midrule
Total & & \LlmComboTotalYesPersonalPct\% & \LlmComboTotalYesDetailsPct\% & \LlmComboTotalYesAppPct\% & \LlmComboTotalYesOtherPct\% & \LlmComboTotalNoPersonalPct\% & \LlmComboTotalNoDetailsPct\% & \LlmComboTotalNoAppPct\% & \LlmComboTotalNoOtherPct\% \\
\bottomrule
\end{tabular}
\end{adjustbox}
\end{table}

\paragraph{User Feedback}
We collected feedback from users on \nodyn{1,298} decisions and the corresponding explanation created by their personalized LLM ($P_{\text{4o}}$), the results are shown in Table~\ref{tab:llm_agreement_by_initial}. Overall,  \dyn{\LlmOverallYesRate}\% of the feedback indicates agreement with the LLM decision, \dyn{\LlmOverallNoRate}\% disagreement, and \dyn{\LlmOverallNotSureRate\%} was reported as unsure. This indicates a generally good reception of LLM decisions and explanations. 
Looking at the feedback on tasks where the LLM agreed with the initial user decision, \dyn{\LlmRateYesInitialagreed\%} of the feedback was positive. 
Feedback on tasks where the user and the LLM initially disagreed is positive in \dyn{\LlmRateYesInitialallowvsonce\%} of cases when the disagreement was on whether it should be \texttt{allow} or \texttt{once}, and positive in \dyn{\LlmRateYesInitialdisagreed\%} of cases when there was a disagreement on whether to \texttt{allow} or \texttt{deny}. 
This indicates that the LLM's reasoning aligned with user values or offered relevant insights, even if the initial decision differed. 
However, \dyn{\LlmRateNoInitialdisagreed\%} of users disagreed with the LLM decision, suggesting that there are gaps in capturing individual preferences.
Table~\ref{tab:llm_agreement_mismatch_split} further splits the \nodyn{611} cases where LLM and user initially disagreed based on the initial user decision. 
Users are more inclined to give positive feedback when the LLM is more restrictive than they were (\dyn{\LlmMismatchYesRateUserAllow\%}) than when it is more permissive (\dyn{\LlmMismatchYesRateUserDeny\%}), indicating value in a system nudging users towards safer behavior and relativizing the increase of functionality violations observed in Table~\ref{tab:personalized_llm_overview}.

The reasons users provided for their (dis-)agreement with LLM decisions are reported in Table~\ref{tab:llm_agreement_reasons_by_combo}. 
When LLM and user initially agreed, the most common reason for positive feedback was that the LLM considered the user's personal preferences---highlighting the importance that users put on personalization---followed by the LLM considering the app's functionality.
In cases of initial disagreement, the LLM considering app details was the most common reason for positive feedback, indicating that users might not have been aware of relevant information when making their decision, while \textit{not} considering personal preferences was the most common negative feedback, indicating gaps in user intent, what they expressed in their privacy statement, and the LLM interpretation of it.

\begin{table}
    \centering
    \caption{\textbf{Adjusted agreement scores for $P_{\text{4o}}$.} Scores are calculated based on initial agreement and the share of ``correct'' LLM decisions (\checkmark) in disagreement cases based on the \expert and user feedback. All values except N in \%. }
    \label{tab:personalized_agreement_by_category}
    \footnotesize
    \begin{tabular}{l r r r r r r r r | r}
    \toprule
    \multicolumn{3}{c}{} & \multicolumn{2}{c}{Expert} & \multicolumn{2}{c}{Feedback}\\
    \cmidrule(lr){4-5} \cmidrule(l){6-7}
    Category & N &  Agree & \checkmark & Adj. & \checkmark & Adj. \\
    \midrule
    No scenario & 10,472 & 70.98 &  -- & -- &  54.87 & 86.90\\
    \midrule
    \Discretion & 2,308 & 65.94 &  -- & -- & 41.14 & 79.95 \\
    \Essential & 921 &  91.42 &   88.61 & 99.02 &  36.00 &  94.50 \\
    \Critical & 981 &  55.25 &   85.42 & 93.48 & 52.74 &  78.85\\
    \bottomrule
    \end{tabular}
\end{table}

\paragraph{Agreement with \Baseline} 
In general, $P_{\text{4o}}$ also aligns well with the expected decision for \baseline tasks: 
\dyn{\PersonalizedVsExpertAccuracyAllow\%} agreement for \essential and \dyn{\PersonalizedVsExpertAccuracyDeny\%} for \critical tasks.
We further look into the \dyn{\MismatchExpertPersTotal} cases where user and LLM decision differed. 
Here, $P_{\text{4o}}$ was aligned with the \baseline \dyn{\MismatchExpertPersAccuracy\%} of the time (\dyn{\MismatchExpertPersAccuracyAllow\%} for \dyn{\MismatchExpertPersTotalAllow} \essential tasks, \dyn{\MismatchExpertPersAccuracyDeny\%} for \dyn{\MismatchExpertPersTotalDeny} \critical tasks), 
showing that the LLM's decision was more in line with the \expert than the user's. However, there are a few decisions where the LLM diverged both from the \expert and the user decision, which is especially concerning for \critical tasks.

\paragraph{Overall Performance} 
To quantify the quality of personalized decisions, we calculate two \textit{adjusted scores} that take into account the initial agreement rate and then the \expert and the user feedback. As feedback was collected for $P_{\text{4o}}$ only and does not cover all decisions, we assume for this calculation that it generalizes to the full $P_{\text{4o}}$ decision set. We calculate the adjusted scores by summing the agreement percentage with the disagreements corrected by a factor of the ``correct'' choices that incorporate the baseline or user feedback.
We report these scores in Table~\ref{tab:personalized_agreement_by_category}.
For example, in no-scenario tasks, $P_{\text{4o}}$ and the initial user decision agreed in \nodyn{70.98\%} of cases. 
We gathered feedback on \nodyn{195} decisions with initial disagreement, which indicated preference for the LLM decision in \nodyn{107 (54.87\%)} of the cases, resulting in an adjusted score of \nodyn{86.90\%}. The adjusted scores are promising: ranging from \nodyn{78.85\%} to \nodyn{99.02\%}, they indicate that personalized LLMs can make decisions that reflect user preferences while also considering circumstantial and security factors.

\subsection{RQ3: Which setup and input factors lead to reasonable LLM decisions?} \label{sec:results_configandsetup}

So far, we have seen that LLMs are capable of making reasonable access control decisions and that it is possible to personalize these decisions to better agree with individual user preferences---with caveats: 
performance varies significantly across users, and the LLM is occasionally more permissive than the \baseline or user.
We therefore investigate which factors can help make such a system practical in spite of these limitations: (1) user input properties and modality, and (2) system configuration choices, including model selection, iterative refinement, and confidence-based deferral,  which can reduce errors while keeping user burden low.

\subsubsection{User Input}
We test H3.1, \textit{a longer and contextually relevant statement of preferences leads to better personalization of LLM decisions} by looking at the correlation of the agreement and length of the statement, and the influence of different input types.

\begin{figure}[t]
    \centering
    \includegraphics[width=\linewidth]{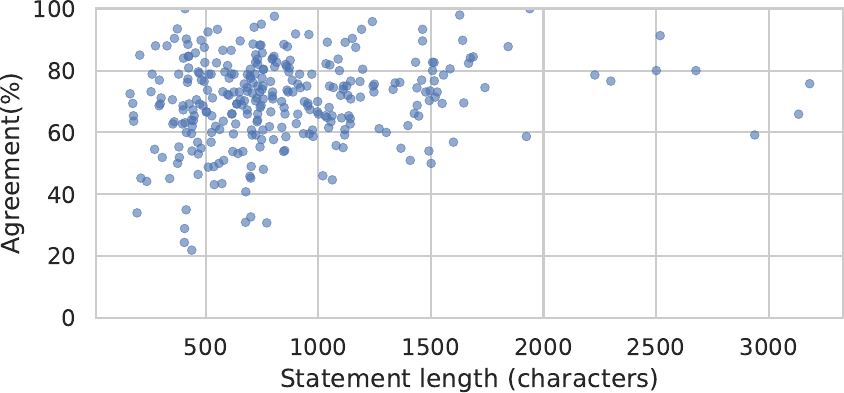}
    \caption{\textbf{Privacy statement length} versus $P_{\text{4o}}$ agreement.}
    \label{fig:input_length_correlation}
    \Description{Scatter plot showing the relationship between the character count of user privacy statements and the agreement rate of the personalized GPT-4o model with individual users.}
\end{figure}

\paragraph{Length of User Input}
We compare the length of a user's privacy statement (measured in characters) and their agreement with $P_{\text{4o}}$ (cf. Figure~\ref{fig:input_length_correlation}).
We observe a positive Pearson correlation ($r(\dyn{\PersInputLenCorrDF}) = \dyn{\PersInputLenCorrPearsonR}$, $p = \nodyn{0.001}$), suggesting that longer user input slightly helps in making better aligned decisions. 
However, we also observe that good alignment can already be achieved with short statements. This is especially true for users with more extreme views---stating a deny-or-allow-all policy can be done in relatively few words. 
Furthermore, longer (shorter) input might indicate higher (lower) engagement during data collection, which could influence the results.

\paragraph{Type of User Input} 
Recall from Section~\ref{sec:casestudy.groups} that participants were exposed to different questions (high-level or phone-focused) and input modes (free text or a conversational LLM) to provide their privacy statements.
A two-way ANOVA revealed no significant main effects for question focus ($F(\dyn{\AnovaGroupDF}, \dyn{\AnovaResidualDF}) = \dyn{\AnovaGroupF}$, $p = \dyn{\AnovaGroupP}$) or input mode ($F(\dyn{\AnovaStyleDF}, \dyn{\AnovaResidualDF}) = \dyn{\AnovaStyleF}$, $p = \dyn{\AnovaStyleP}$) on per-user agreement with $\text{P}_{\text{4o}}$.
While there is a statistically significant interaction effect ($F(\dyn{\AnovaInteractionDF}, \dyn{\AnovaResidualDF}) = \dyn{\AnovaInteractionF}$, $p = \dyn{\AnovaInteractionP}$), the practical differences are modest.
This supports the collection of ``high-level'' privacy statements that may transfer to other contexts, such as agentic systems.

\begin{figure}[t]
    \centering
    \includegraphics[width=\linewidth]{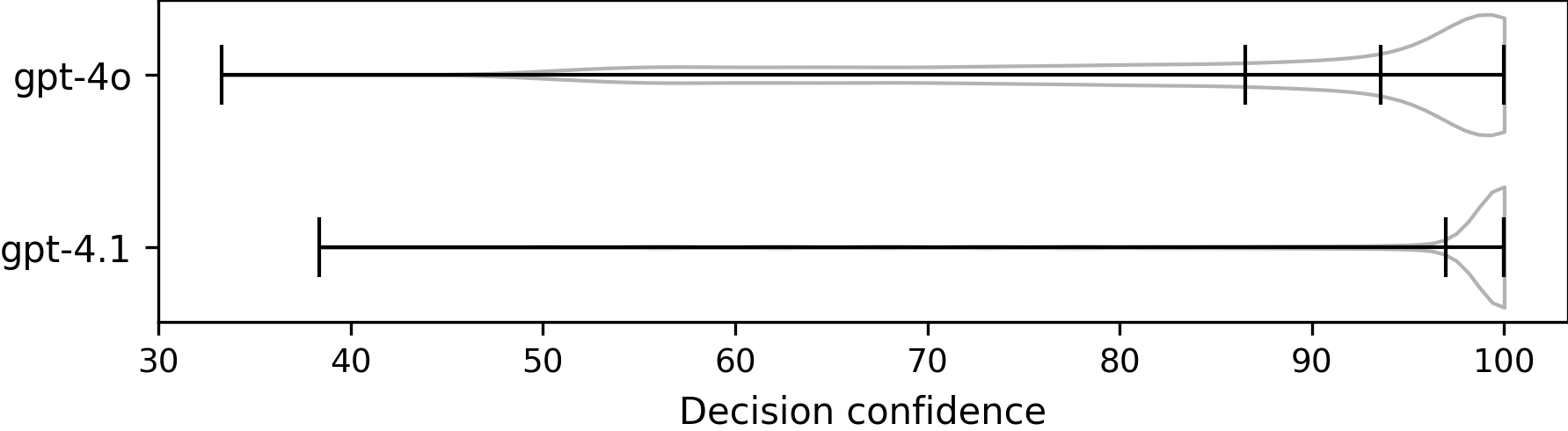}
    \caption{\textbf{Distribution of confidence} for GPT-4o and 4.1.}
    \label{fig:confidence_score_distributions}
    \Description{Violin plots showing the distribution of LLM confidence scores for allow and deny decisions across GPT-4o and GPT-4.1. }
\end{figure}

\paragraph{User Consistency}
In a qualitative inspection of the 10 users with the highest positive and the 10 users with the highest negative impact of personalization, we observe the following patterns: for the users with positive change in agreement, we found very strong statements as part of their privacy statement, such as \textit{``I grant all permissions an app wants/needs. I would rather get the full functionality of an app/service, as I don't really have any concerns about my persona data [sic]''}, or \textit{``If I'm using an app for directions I will allow permissions and when I'm done I close it out and disable permissions.''}, which also reflect their choices, i.e., very high allow or deny rates.
Among users with negative change, we observe two patterns: (1) a strong statement and a contradictory decision-making behavior, emerging in both directions, with users making very permissive statements, but are restrictive in their decision-making, and the opposite, and (2) a balanced statement and mixed decision-making.
This inspection is illustrative: a systematic analysis of the alignment of statement and behavior remains future work.  

To validate that study decisions reflect real-world behavior, we used data from the optional part of the study, in which participants reported permissions they had granted to apps. For \dyn{\PhoneCfgCompareAccuracy\%} of the \dyn{\PhoneCfgCompareTotal} app-permission pairs for which a corresponding no-scenario study decision existed, participants also chose \textit{allow} in the study, providing evidence for the external validity of our study, though agreement varied by permission type (Camera: \dyn{\PhoneCfgCameraAccuracy\%}, Microphone: \dyn{\PhoneCfgMicrophoneAccuracy\%}, Photos: \dyn{\PhoneCfgPhotosAccuracy\%}, Calendar: \dyn{\PhoneCfgCalendarAccuracy\%}).

\subsubsection{System Configuration}
Finally, we turn to our hypothesis 3.2, that \textit{LLM-based access control can be configured to minimize false positives while maintaining utility and usability}.

\paragraph{Choice of LLM} Model choice matters---each comes with its own biases. E.g., the agreement with the majority votes and expert baseline varies between different models using the same prompt (cf. Table~\ref{tab:user_decision_overview}). From our experiment with personalized models, there are also indications that this bias ``bleeds'' into personalized decisions. One can see this, for example, when one looks at the 31 no-scenario tasks for which more than 2/3 of users choose deny. $G_{\text{4o}}$ decided on \texttt{allow} for 3 out of these scenarios, and \texttt{deny} for the other 28. For the first 3 scenarios, $P_{\text{4o}}$ had an average deny rate of \nodyn{34.8\%}, while the average deny rate for the other 28 tasks was \nodyn{91.4\%}. We did not systematically analyze this effect, but deem it an important consideration for system designers.

Further, the confidence scores produced by different LLMs show varying degrees of correlation with the user consensus, indicating that some models may provide more reliable confidence estimates than others (if it is even possible to get them at all, e.g., no logprobs for GPT-5). For the two models that provide logprobs (GPT-4o and GPT-4.1), Figure~\ref{fig:confidence_score_distributions} shows the distribution of confidence scores. GPT-4o produces a more distributed set of confidence scores with a mean of \nodyn{86.5} (std: \nodyn{15.3}), while GPT-4.1 skews towards higher confidence scores with a mean of \nodyn{97.0} (std: \nodyn{8.9}). This indicates that GPT-4o's scores are more reflective of actual uncertainty. 

\paragraph{Influence of confidence thresholds}
In our analysis so far, we have considered all LLM decisions equally. However, in a practical setup, it might be desirable to only accept LLM decisions that are made with high confidence, and to fall back to alternative mechanisms (e.g., user input) when the LLM is uncertain.
To investigate this, we analyze the impact of applying different confidence thresholds. A higher confidence threshold means that a lower number of decisions are considered, which could lead to higher user fatigue or annoyance, but it could improve the overall accuracy and security of the system. It might be desirable to have different thresholds for \texttt{allow} and \texttt{deny} decisions, as the cost of misclassifying these two types of decisions can be different. 

\begin{figure*}[t]
  \centering
  \begin{minipage}[t]{0.24\textwidth}
    \centering
    \includegraphics[width=\linewidth]{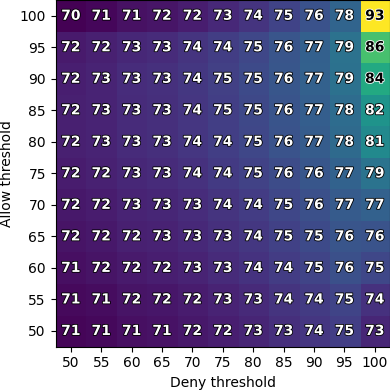}
    \\[3pt]
    \small Agreement Rate
  \end{minipage}%
  \hfill
    \begin{minipage}[t]{0.24\textwidth}
    \centering
    \includegraphics[width=\linewidth]{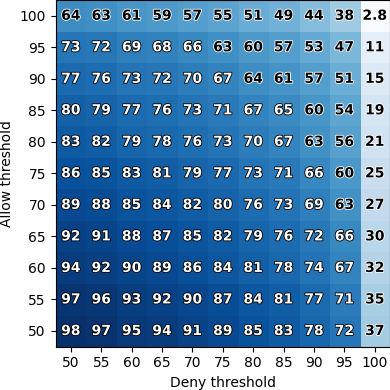}
    \\[3pt]
    \small Coverage
  \end{minipage}%
  \hfill
  \begin{minipage}[t]{0.24\textwidth}
    \centering
    \includegraphics[width=\linewidth]{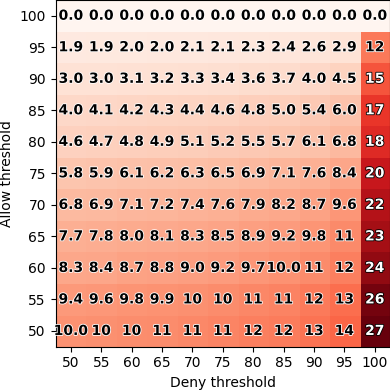}
    \\[3pt]
    \small Security violations
  \end{minipage}%
  \hfill
  \begin{minipage}[t]{0.24\textwidth}
    \centering
    \includegraphics[width=\linewidth]{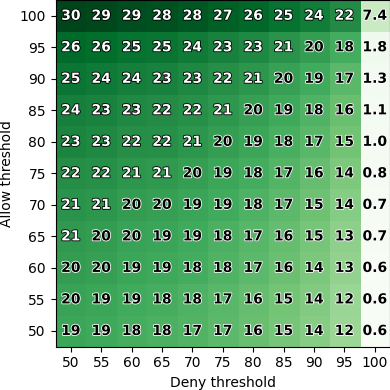}
    \\[3pt]
    \small Functionality violations
  \end{minipage}%

  \caption{\textbf{Heatmaps of $\text{P}_{\text{4o}}$ decision performance} for different confidence thresholds. All values are percentages.}
  \label{fig:gpt4o-pv7-heatmaps}
  \Description{Four heatmaps showing personalized GPT-4o decision performance at varying confidence thresholds: agreement rate, coverage, security violations, and functionality violations. Rows are confidence thresholds, columns are task types.}
\end{figure*}

Figure \ref{fig:gpt4o-pv7-heatmaps} shows the accuracy, coverage, security violations, and functionality violations over all $\text{P}_{\text{4o}}$ decisions. E.g., for our dataset, when setting a confidence threshold of 100\% for allow decisions and 50\% for deny decisions, we achieve an overall accuracy of \nodyn{70.4\%}, but make decisions for only \nodyn{64.3\%} of all tasks. In this setting, we observe \nodyn{0} decisions where the LLM would allow access even though the user would not, but of all decisions made, the LLM chose to deny while the user chose to allow in \nodyn{30\%} of cases. Based on such results, system designers can make informed choices about the trade-offs involved in setting confidence thresholds. For security-sensitive deployments, we recommend prioritizing a high threshold for \texttt{allow} and a moderate one for \texttt{deny}: this minimizes security violations while still automating the majority of decisions, deferring the remainder to the user.

\paragraph{Iterative Refinement of User Preferences}
A realistic deployment of an LLM-based access control system would resemble an interactive, feedback-driven design: the model progressively refines its decisions as users provide explicit feedback or override individual decisions. We take a first step toward evaluating this by simulating it via in-context learning, providing subsets of user decisions and feedback to the LLM.
To assess how example type and quantity influence performance, we randomly sampled eight tasks per user (four with a scenario, four without) as in-context feedback, and evaluated the remaining tasks. We compare the model's performance when given only scenario tasks, only no-scenario tasks, two scenario and two no-scenario tasks, or all eight together. 
Despite random sampling, the example distributions were generally balanced, with slightly more allows in scenario tasks (58\% allow samples) and more denies in no-scenario (43\% allow) tasks. 
Overall agreement remained stable, decreasing slightly (\nodyn{-1pp}). However, incorporating user decisions made the model more conservative: security violations decreased by \nodyn{5 to 7pp}, while functionality violations increased by \nodyn{7 to 8pp}.
Reflecting skewed/simpler user decision distributions, the largest improvements occurred at the extremes, when the model received only allow or only deny examples. 

From the free-text feedback collected on LLM decisions (cf. Section \ref{sec:setup.study_steps}), we identified 71 users who provided concrete, actionable feedback by pre-classifying responses with an LLM and then manually checking them. We evaluate two settings: using this feedback alone and combining it with the eight examples from the previous setting. Compared to the personalized baseline, agreement improved by \nodyn{1.0 and 1.4pp}, respectively, but security violations decreased (\nodyn{6.6pp and 9.0pp}), while functionality violations increased (\nodyn{5.6pp and 7.6pp}), indicating that feedback reinforced cautious behavior.
Together, these results provide an initial empirical basis for interactive, feedback-driven access control designs, where user corrections progressively align the system with individual preferences.

\section{Discussion}
\subsection{Main Insights}

\paragraph{Generic LLMs make reasonable access control decisions}
LLMs agreed with the majority vote of users for \nodyn{75\%}--\nodyn{89\%} of our study tasks, demonstrating that a non-personalized LLM can take reasonable decisions for most common users based on its trained knowledge about the system and its apps, and its capability to parse scenario context. 
In clear cases, different models often aligned with the \baseline decision, correctly granting or denying permissions as expected. However, there is a gap for some models for \critical tasks, indicating a need to reinforce model behavior accordingly. Overall, LLMs still took safer
decisions than the user majority.

\paragraph{Personalized LLMs do not fit all users}
The value of personalization depends heavily on the user: the overall improvement is modest (+2--5\%), but for users who can articulate clear preferences or hold extreme views, gains can exceed 50pp, while users whose statements contradict their behavior may actually see agreement decrease.
In rare cases, personalization can introduce security violations: a customized LLM may grant access that both the \baseline and the generic model would deny, particularly when aligning with users who stated very permissive preferences.

\paragraph{LLM decisions can be safer or preferred by users}
We carefully analyzed disagreements between LLM and users with the assumption that the user's decision is not always the right one.
When a personalized LLM disagrees with a user's initial choice, its explanation can be persuasive: around 49\% of users who initially made a different decision ultimately agreed with the LLM after reading its reasoning. 
This effect was particularly pronounced when the LLM recommended a more restrictive choice than the user, with around 60\% of users agreeing with it. This is practically important: safety and user acceptance seem to align---a conservative-leaning system reduces security violations while remaining acceptable to users.
Overall, by combining agreement rates with feedback on disagreements and expected \baseline decisions, the LLM made ``correct'' choices between 79\% and 99\% of the time. 

\paragraph{LLMs can be configured to reduce access control decision burden}
Confidence-based deferral further sharpens this picture: 
For GPT-4o, we observed that decision token probabilities correlated with the strength of user consensus 
and that agreement with individual user decisions increases with higher decision confidence. 
This provides a robust mechanism to decide whether to make an automated decision or defer to the user. For a security-focused system, setting a high confidence threshold for \texttt{allow} decisions and a moderate one for \texttt{deny} decisions can help to balance safety with utility: our personalized GPT-4o model would be able to autonomously handle \nodyn{64}\% of cases while making no security-relevant errors on our data.

Together, our findings suggest a practical deployment strategy: a generic LLM handles clear-cut cases well, personalization benefits users with well-specified preferences, and confidence-based deferral gracefully manages uncertainty.
Two complementary approaches to reach well-specified preferences seem like an interesting avenue for further investigation to us: iterative refinement, where the system incorporates user corrections over time, and simply asking users whether they want natural-language-based personalization---those who opt in are potentially more likely to hold strong, consistent views and to be able to express them clearly.

\subsection{Limitations} \label{sec:limitations}
Our work and its analysis provide valuable insights into the opportunities and challenges of using LLM-assisted access control, but they are also subject to several limitations that point to important directions for future work.

\paragraph{Broader Applicability}
We examine one access control system, mobile app permission management. 
How our insights generalize to different systems needs additional investigation: personalized LLM-based access control is an interesting option for any domain where access decisions depend on user preferences and an LLM can provide relevant domain knowledge. 
Agentic systems such as OpenClaw~\cite{openclaw} are a compelling case: as agents interact with external services and resources,  a dedicated LLM access control guard can act on the user's behalf  using preference statements. 
Compared to mobile permissions, such a setting offers richer per-decision context (full task descriptions rather than only permission request context), which might improve LLM-user agreement. 
Additionally, the higher decision frequency in agentic settings makes the trade-off between automation and user burden and the confidence-based deferral mechanism particularly important. 
However, the more specialized or novel a domain, the less prior knowledge an LLM can draw on. Our study benefits from LLMs' knowledge of the Android ecosystem, but niche systems (e.g., proprietary smart home platforms, enterprise systems) may lack it. In such settings, providing the LLM with rich, domain-specific information so that it can apply its security reasoning becomes increasingly important and remains an open problem.
The personalization challenges we observed, such as preference--behavior inconsistencies and statement quality, are likely to carry over to other use cases.

\paragraph{Design and Generalizability}
Our data collection, while carefully designed, took place in an experimental setting. This could influence participant behavior in two ways: (1) They could have been more focused on the decision than they would be during a real permission request, or (2) decisions were hypothetical and did not have any consequences, making participants potentially more careless. Our data collection is limited to English-speaking participants and residents of the UK and the US; our findings might not generalize to other languages or cultures with different privacy norms and expectations.
As user feedback was collected after reading the LLM's justification, by design, agreement rates reflect users finding the reasoning convincing. However, we cannot distinguish genuine correctness from mere persuasiveness.

\paragraph{Binational confound} 
We conducted exploratory analyses comparing US (n=\dyn{\USPerUserNUsers}) and UK (n=\dyn{\UKPerUserNUsers}) participants across key metrics. Slight numerical differences, e.g., in deny rates (US: \dyn{\USOverallDenyRate}\%, UK: \dyn{\UKOverallDenyRate}\%, two-sample z-test $z=\dyn{\CountryDenyRateTestZ}$, $p=\dyn{\CountryDenyRateTestP}$) and personalized LLM agreement (US: 69.23\%, UK: 71.72\%, independent samples t-test $t(305)=\dyn{\CountryPerUserTestT}$, $ p=\dyn{\CountryPerUserTestP}$) indicate only marginal cross-country variation. Therefore, we pooled data from both countries in our main analyses to increase statistical power.

\paragraph{Threat model}
We assume a threat model in which apps are not attacking the system by employing methods aimed at LLM inference, e.g., by poisoning external knowledge sources an LLM might use. Future work should investigate the resilience of the system and defense mechanisms against such attacks.

\paragraph{Statement Quality and System Design}
The effectiveness of personalization depends critically on the quality of privacy statements. This points to two directions for future work: (1) investigating how to better support users in providing accurate, informative statements and (2) designing systems that detect when personalization is potentially harmful and fall back to safer defaults.

\paragraph{LLM limitations}
LLMs are stochastic processes that can produce inconsistent and unpredictable outputs~\cite{atil2025nondeterminismdeterministicllmsettings}. This must be taken into consideration and compared against the risk of alternatives (e.g., decision fatigue leading a user to make arbitrary decisions). Further, one needs to consider the cost and delay introduced by LLM inference. However, especially if LLM access control is embedded into agentic systems, it should introduce a modest relative overhead.

\section{Related Work}

\paragraph{Natural-language Access Control}
Due to the need for complex access control policies, several works have explored using natural language and NLP to synthesize them.
Early works~\cite{xiao2012automated,shi2011controlled} use NLP techniques to extract access-control rules from natural-language software requirements and map them to XACML-style policies.
This opens new research directions that enhance NLP to synthesize more robust policies~\cite{de2009using,narouei2017identification,nguyen2015automatic,planas2023modeling}.
Due to recent advancements in LLMs, studies have begun replacing traditional NLP techniques with them.
\cite{yang2025extraction} proposes a two-step pipeline that uses LLMs and knowledge distillation to convert natural-language access control policies (NLACPs) into machine-enforceable ABAC rules, with evaluation on realistic policy texts. 
Similarly, other works~\cite{sonune2025lmn,cheng2025say,lawal2024translating,yang2025extraction,subramaniam2025deploiapplyingnl2sqlsynthesize} use LLMs to synthesize access control policies from natural language.
Apart from LLMs, deep learning~\cite{shan2024automatic, 10.1145/3508398.3511497} methods are also used to mine access control policies from both real and synthetic user traces.  In contrast to the approach we investigate, these works rely on an existing specification of policies. Wu et al.~\cite{wu2025towards} demonstrated that user permission decisions for agents are predictable based on contextual factors identified through a vignette study. Our work is complementary: while they build a predictive ML-based assistant, we evaluate the use of LLMs as an intuitive reasoning bridge that interprets high-level, natural-language privacy sentiments to make aligned decisions without explicit feature engineering.

\paragraph{LLMs and User Behavior}
LLMs learning and emulating human preferences is central to this paper.
Several studies~\cite{tjuatja2024llms,geva2025llms,chen2024humans,li2024dissecting} reveal that there are gaps between the LLMs' behavior and humans in several survey-style studies.
In contrast, other works~\cite{ji2023beavertails,song2024preference,jiang2024survey,goli2023can} show that LLMs can learn human preferences.
Other work investigates LLM capability to track sensitive information in increasingly complex setups~\cite{canllmkeepsecret} and finds that LLMs are likely to overshare in comparison to a human baseline. \cite{llmprivacyjudge}~compares privacy evaluation of LLMs against user baselines and finds that inter-human agreement is low, but that LLMs capture the overall human perspective.

\paragraph{Access Control on Phones}
Taintdroid~\cite{enck2014taintdroid} instruments Android’s runtime to perform real-time, system-wide dynamic taint tracking of sensitive sources  to detect when tainted data leaves the device.
Apex~\cite{nauman2010apex} extends Android’s permission model with a policy-enforcement framework that lets users specify per-app runtime constraints.
FlaskDroid~\cite{bugiel2013flexible} brings a Flask-style~\cite{spencer1999flask} MAC architecture to Android, aligning middleware and kernel enforcement to support a range of fine-grained policies. 
SEApp~\cite{rossi2021seapp} proposes a practical MAC approach for Android apps that enforces policies at app granularity with compatibility and deployability in mind.
Finedroid~\cite{zhang2015finedroid} implements context-sensitive, system-wide permission enforcement that adapts permission decisions to runtime context to reduce over-privilege while preserving app functionality.
All of these approaches still introduce cognitive load on the user. 

\paragraph{Access Control Studies}
\cite{felt2011android}~conducts an extensive empirical study of Android apps showing widespread over-privilege and mapping of API calls permissions. It also shows that app developers intended to follow the least privilege principle, but sometimes fail due to insufficient API documentation.
\cite{felt2012android}~conducted an online and lab study to measure whether users notice and understand install-time permissions and how this affects install decisions; it found low comprehension and attention.
Wijesekera et al.~\cite{wijesekera2015android} show through a 36-participant survey that permission requests should appear during runtime, rather than Android’s previous install-time approval approach. 
PrivaDroid~\cite{cao2021androidlargescalestudy} conducts a large survey exploring how users understand permissions, what they expect apps will do, and mismatch patterns that produce perceived risk. Turtle Guard~\cite{turtleguard} focuses on auditing and feedback mechanisms to improve automated permission managers.

\paragraph{Predicting Privacy Preferences}
Beyond studying how users perceive and respond to permission requests, prior work has investigated how to model individual privacy preferences.
Wijesekera et al.~\cite{wijesekera2018contextualizing} show that incorporating contextual signals substantially improves the prediction of privacy decisions.
\cite{smullen2020best} investigates ML-driven recommender systems to mitigate the trade-off between user burden and the accurate configuration of mobile app privacy preferences. 
Liu et al.~propose personalized privacy assistants that recommend permission decisions based on learned user profiles~\cite{DBLP:conf/www/LiuLS14, DBLP:conf/soups/0017ASAZSAA16}. These systems demonstrate the feasibility of modelling user-specific privacy preferences, but typically rely on historical user decisions, predefined features rather than open-ended natural-language expressions of privacy intent, and require dedicated training.

\section{Conclusion}

We investigated whether LLMs are capable of making decisions on mobile app permission requests to relieve users of the related cognitive burden. 
We collected a dataset of personal preferences and access-control decisions against which to compare LLM decisions.
Our results are encouraging: we observed a good alignment between generic LLMs and the human majority vote; personalized LLM decisions match the user decision \gptAoPersonalizedVsUsersOverallAcc\% of the time. 
Taking into account users' feedback on disagreements and \baseline tasks, we measured an LLM success rate of 79\% to 99\%. 
However, the impact varies per user and in rare cases, an LLM made less safe decisions than a user.
Combined, our results show that LLMs can help reduce user burden in making access control decisions, especially when deferring uncertain decisions to users.
Future work should explore how to help users express their privacy statements to increase the benefits of personalization and test the concept of LLMs for access control in different settings (e.g., managing access control in a corporation)
and understand how to tune an LLM-assisted access control system to further protect users in sensitive scenarios.

\bibliographystyle{IEEEtranS}
\bibliography{references}

\begin{thebibliography}{10}
\providecommand{\url}[1]{#1}
\csname url@samestyle\endcsname
\providecommand{\newblock}{\relax}
\providecommand{\bibinfo}[2]{#2}
\providecommand{\BIBentrySTDinterwordspacing}{\spaceskip=0pt\relax}
\providecommand{\BIBentryALTinterwordstretchfactor}{4}
\providecommand{\BIBentryALTinterwordspacing}{\spaceskip=\fontdimen2\font plus
\BIBentryALTinterwordstretchfactor\fontdimen3\font minus \fontdimen4\font\relax}
\providecommand{\BIBforeignlanguage}[2]{{%
\expandafter\ifx\csname l@#1\endcsname\relax
\typeout{** WARNING: IEEEtranS.bst: No hyphenation pattern has been}%
\typeout{** loaded for the language `#1'. Using the pattern for}%
\typeout{** the default language instead.}%
\else
\language=\csname l@#1\endcsname
\fi
#2}}
\providecommand{\BIBdecl}{\relax}
\BIBdecl

\bibitem{openclaw}
``{OpenClaw},'' \url{https://openclaw.ai/}, accessed: 2026-04-12.

\bibitem{prolific}
``Prolific,'' \url{https://www.prolific.com/}.

\bibitem{acquisti2015privacy}
A.~Acquisti, L.~Brandimarte, and G.~Loewenstein, ``Privacy and human behavior in the age of information,'' \emph{Science}, vol. 347, no. 6221, pp. 509--514, 2015.

\bibitem{AcquistiG05}
\BIBentryALTinterwordspacing
A.~Acquisti and J.~Grossklags, ``Privacy and rationality in individual decision making,'' \emph{{IEEE} Secur. Priv.}, vol.~3, no.~1, pp. 26--33, 2005. [Online]. Available: \url{https://doi.org/10.1109/MSP.2005.22}
\BIBentrySTDinterwordspacing

\bibitem{ALBERT201049}
\BIBentryALTinterwordspacing
B.~Albert, T.~Tullis, and D.~Tedesco, ``Chapter 3 - designing the study,'' in \emph{Beyond the Usability Lab}, B.~Albert, T.~Tullis, and D.~Tedesco, Eds.\hskip 1em plus 0.5em minus 0.4em\relax Boston: Morgan Kaufmann, 2010, pp. 49--92. [Online]. Available: \url{https://www.sciencedirect.com/science/article/pii/B978012374892800003X}
\BIBentrySTDinterwordspacing

\bibitem{atil2025nondeterminismdeterministicllmsettings}
\BIBentryALTinterwordspacing
B.~Atil, S.~Aykent, A.~Chittams, L.~Fu, R.~J. Passonneau, E.~Radcliffe, G.~R. Rajagopal, A.~Sloan, T.~Tudrej, F.~Ture, Z.~Wu, L.~Xu, and B.~Baldwin, ``Non-determinism of "deterministic" llm settings,'' 2025. [Online]. Available: \url{https://arxiv.org/abs/2408.04667}
\BIBentrySTDinterwordspacing

\bibitem{bentegeac2025token}
R.~Bentegeac, B.~Le~Guellec, G.~Kuchcinski, P.~Amouyel, and A.~Hamroun, ``Token probabilities to mitigate large language models overconfidence in answering medical questions: Quantitative study,'' \emph{Journal of medical Internet research}, vol.~27, p. e64348, 2025.

\bibitem{bugiel2013flexible}
S.~Bugiel, S.~Heuser, and A.-R. Sadeghi, ``Flexible and fine-grained mandatory access control on android for diverse security and privacy policies,'' in \emph{22nd USENIX Security Symposium (USENIX Security 13)}, 2013, pp. 131--146.

\bibitem{cao2021androidlargescalestudy}
\BIBentryALTinterwordspacing
W.~Cao, C.~Xia, S.~T. Peddinti, D.~Lie, N.~Taft, and L.~M. Austin, ``A large scale study of user behavior, expectations and engagement with android permissions,'' in \emph{30th {USENIX} Security Symposium, {USENIX} Security 2021, August 11-13, 2021}, M.~D. Bailey and R.~Greenstadt, Eds.\hskip 1em plus 0.5em minus 0.4em\relax {USENIX} Association, 2021, pp. 803--820. [Online]. Available: \url{https://www.usenix.org/conference/usenixsecurity21/presentation/cao-weicheng}
\BIBentrySTDinterwordspacing

\bibitem{chen2024humans}
G.~H. Chen, S.~Chen, Z.~Liu, F.~Jiang, and B.~Wang, ``Humans or llms as the judge? a study on judgement biases,'' \emph{arXiv preprint arXiv:2402.10669}, 2024.

\bibitem{cheng2025say}
Y.~Cheng, M.~Xu, Y.~Zhang, K.~Li, H.~Wu, Y.~Zhang, S.~Guo, W.~Qiu, D.~Yu, and X.~Cheng, ``Say what you mean: Natural language access control with large language models for internet of things,'' \emph{arXiv preprint arXiv:2505.23835}, 2025.

\bibitem{de2009using}
\BIBentryALTinterwordspacing
J.~L.~D. Coi, P.~K{\"{a}}rger, D.~Olmedilla, and S.~Zerr, ``Using natural language policies for privacy control in social platforms,'' in \emph{Proceedings of the {ESWC2009} Workshop on Trust and Privacy on the Social and Semantic Web {(SPOT2009)} Heraklion, Greece, June 1, 2009}, ser. {CEUR} Workshop Proceedings, M.~Hausenblas, P.~K{\"{a}}rger, D.~Olmedilla, A.~Passant, and A.~Polleres, Eds., vol. 447.\hskip 1em plus 0.5em minus 0.4em\relax CEUR-WS.org, 2009. [Online]. Available: \url{https://ceur-ws.org/Vol-447/paper4.pdf}
\BIBentrySTDinterwordspacing

\bibitem{bias_llms}
\BIBentryALTinterwordspacing
S.~Dai, C.~Xu, S.~Xu, L.~Pang, Z.~Dong, and J.~Xu, ``Bias and unfairness in information retrieval systems: New challenges in the {LLM} era,'' in \emph{Proceedings of the 30th {ACM} {SIGKDD} Conference on Knowledge Discovery and Data Mining, {KDD} 2024, Barcelona, Spain, August 25-29, 2024}, R.~Baeza{-}Yates and F.~Bonchi, Eds.\hskip 1em plus 0.5em minus 0.4em\relax {ACM}, 2024, pp. 6437--6447. [Online]. Available: \url{https://doi.org/10.1145/3637528.3671458}
\BIBentrySTDinterwordspacing

\bibitem{dong2024survey}
Q.~Dong, L.~Li, D.~Dai, C.~Zheng, J.~Ma, R.~Li, H.~Xia, J.~Xu, Z.~Wu, B.~Chang \emph{et~al.}, ``A survey on in-context learning,'' in \emph{Proceedings of the 2024 conference on empirical methods in natural language processing}, 2024, pp. 1107--1128.

\bibitem{egelman2013choice}
S.~Egelman, A.~P. Felt, and D.~Wagner, ``Choice architecture and smartphone privacy: There’sa price for that,'' in \emph{The economics of information security and privacy}.\hskip 1em plus 0.5em minus 0.4em\relax Springer, 2013, pp. 211--236.

\bibitem{enck2014taintdroid}
W.~Enck, P.~Gilbert, S.~Han, V.~Tendulkar, B.-G. Chun, L.~P. Cox, J.~Jung, P.~McDaniel, and A.~N. Sheth, ``Taintdroid: an information-flow tracking system for realtime privacy monitoring on smartphones,'' \emph{ACM Transactions on Computer Systems (TOCS)}, vol.~32, no.~2, pp. 1--29, 2014.

\bibitem{fadeeva2024fact}
E.~Fadeeva, A.~Rubashevskii, A.~Shelmanov, S.~Petrakov, H.~Li, H.~Mubarak, E.~Tsymbalov, G.~Kuzmin, A.~Panchenko, T.~Baldwin \emph{et~al.}, ``Fact-checking the output of large language models via token-level uncertainty quantification,'' \emph{arXiv preprint arXiv:2403.04696}, 2024.

\bibitem{fang2010privacy}
L.~Fang and K.~LeFevre, ``Privacy wizards for social networking sites,'' in \emph{Proceedings of the 19th international conference on World wide web}, 2010, pp. 351--360.

\bibitem{felt2011android}
A.~P. Felt, E.~Chin, S.~Hanna, D.~Song, and D.~Wagner, ``Android permissions demystified,'' in \emph{Proceedings of the 18th ACM conference on Computer and communications security}, 2011, pp. 627--638.

\bibitem{felt2012android}
A.~P. Felt, E.~Ha, S.~Egelman, A.~Haney, E.~Chin, and D.~Wagner, ``Android permissions: User attention, comprehension, and behavior,'' in \emph{Proceedings of the eighth symposium on usable privacy and security}, 2012, pp. 1--14.

\bibitem{fisher2000social}
R.~J. Fisher and J.~E. Katz, ``Social-desirability bias and the validity of self-reported values,'' \emph{Psychology \& marketing}, vol.~17, no.~2, pp. 105--120, 2000.

\bibitem{galesic2009effects}
M.~Galesic and M.~Bosnjak, ``Effects of questionnaire length on participation and indicators of response quality in a web survey,'' \emph{Public opinion quarterly}, vol.~73, no.~2, pp. 349--360, 2009.

\bibitem{qualtrics_commitment}
\BIBentryALTinterwordspacing
E.~Geisen, ``\BIBforeignlanguage{en}{Using {Commitment} {Requests} {Instead} of {Attention} {Checks}}.'' [Online]. Available: \url{https://www.qualtrics.com/articles/strategy-research/attention-checks-and-data-quality/}
\BIBentrySTDinterwordspacing

\bibitem{geva2025llms}
T.~Geva, A.~Goldstein, E.~Lary, and C.~Levy, ``Do llms exhibit human-like cognitive biases? a large-scale systematic evaluation,'' \emph{A Large-Scale Systematic Evaluation (September 17, 2025)}, 2025.

\bibitem{goli2023can}
A.~Goli and A.~Singh, ``Can llms capture human preferences?'' \emph{arXiv preprint arXiv:2305.02531}, 2023.

\bibitem{androidPermissionsAndroid}
Google, ``{P}ermissions on {A}ndroid,'' \url{https://developer.android.com/guide/topics/permissions/overview}, [Accessed 07-11-2025].

\bibitem{he2024doespromptformattingimpact}
\BIBentryALTinterwordspacing
J.~He, M.~Rungta, D.~Koleczek, A.~Sekhon, F.~X. Wang, and S.~Hasan, ``Does prompt formatting have any impact on llm performance?'' 2024. [Online]. Available: \url{https://arxiv.org/abs/2411.10541}
\BIBentrySTDinterwordspacing

\bibitem{huang2024large}
T.~Huang, L.~You, N.~Cai, and T.~Huang, ``Large language model firewall for aigc protection with intelligent detection policy,'' in \emph{2024 2nd International Conference On Mobile Internet, Cloud Computing and Information Security (MICCIS)}.\hskip 1em plus 0.5em minus 0.4em\relax IEEE, 2024, pp. 247--252.

\bibitem{ji2023beavertails}
J.~Ji, M.~Liu, J.~Dai, X.~Pan, C.~Zhang, C.~Bian, B.~Chen, R.~Sun, Y.~Wang, and Y.~Yang, ``Beavertails: Towards improved safety alignment of llm via a human-preference dataset,'' \emph{Advances in Neural Information Processing Systems}, vol.~36, pp. 24\,678--24\,704, 2023.

\bibitem{jiang2024survey}
R.~Jiang, K.~Chen, X.~Bai, Z.~He, J.~Li, M.~Yang, T.~Zhao, L.~Nie, and M.~Zhang, ``A survey on human preference learning for large language models,'' \emph{arXiv preprint arXiv:2406.11191}, 2024.

\bibitem{lawal2024translating}
S.~Lawal, X.~Zhao, A.~Rios, R.~Krishnan, and D.~Ferraiolo, ``Translating natural language specifications into access control policies by leveraging large language models,'' in \emph{2024 IEEE 6th International Conference on Trust, Privacy and Security in Intelligent Systems, and Applications (TPS-ISA)}.\hskip 1em plus 0.5em minus 0.4em\relax IEEE, 2024, pp. 361--370.

\bibitem{li2024dissecting}
J.~Li, F.~Zhou, S.~Sun, Y.~Zhang, H.~Zhao, and P.~Liu, ``Dissecting human and llm preferences,'' \emph{arXiv preprint arXiv:2402.11296}, 2024.

\bibitem{DBLP:conf/soups/0017ASAZSAA16}
\BIBentryALTinterwordspacing
B.~Liu, M.~S. Andersen, F.~Schaub, H.~Almuhimedi, S.~Zhang, N.~M. Sadeh, Y.~Agarwal, and A.~Acquisti, ``Follow my recommendations: {A} personalized privacy assistant for mobile app permissions,'' in \emph{Twelfth Symposium on Usable Privacy and Security, {SOUPS} 2016, Denver, CO, USA, June 22-24, 2016}.\hskip 1em plus 0.5em minus 0.4em\relax {USENIX} Association, 2016, pp. 27--41. [Online]. Available: \url{https://www.usenix.org/conference/soups2016/technical-sessions/presentation/liu}
\BIBentrySTDinterwordspacing

\bibitem{DBLP:conf/www/LiuLS14}
\BIBentryALTinterwordspacing
B.~Liu, J.~Lin, and N.~M. Sadeh, ``Reconciling mobile app privacy and usability on smartphones: could user privacy profiles help?'' in \emph{23rd International World Wide Web Conference, {WWW} '14, Seoul, Republic of Korea, April 7-11, 2014}, C.~Chung, A.~Z. Broder, K.~Shim, and T.~Suel, Eds.\hskip 1em plus 0.5em minus 0.4em\relax {ACM}, 2014, pp. 201--212. [Online]. Available: \url{https://doi.org/10.1145/2566486.2568035}
\BIBentrySTDinterwordspacing

\bibitem{mei2025reasoninguncertaintyreasoningmodels}
\BIBentryALTinterwordspacing
Z.~Mei, C.~Zhang, T.~Yin, J.~Lidard, O.~Shorinwa, and A.~Majumdar, ``Reasoning about uncertainty: Do reasoning models know when they don't know?'' 2025. [Online]. Available: \url{https://arxiv.org/abs/2506.18183}
\BIBentrySTDinterwordspacing

\bibitem{llmprivacyjudge}
\BIBentryALTinterwordspacing
S.~Meisenbacher, A.~Klymenko, and F.~Matthes, ``Llm-as-a-judge for privacy evaluation? exploring the alignment of human and {LLM} perceptions of privacy in textual data,'' \emph{CoRR}, vol. abs/2508.12158, 2025. [Online]. Available: \url{https://doi.org/10.48550/arXiv.2508.12158}
\BIBentrySTDinterwordspacing

\bibitem{mhatre2025llm}
A.~Mhatre, N.~Nader, P.~Diehl, and D.~Gupta, ``Llm-guard: Large language model-based detection and repair of bugs and security vulnerabilities in c++ and python,'' \emph{arXiv preprint arXiv:2508.16419}, 2025.

\bibitem{canllmkeepsecret}
\BIBentryALTinterwordspacing
N.~Mireshghallah, H.~Kim, X.~Zhou, Y.~Tsvetkov, M.~Sap, R.~Shokri, and Y.~Choi, ``Can llms keep a secret? testing privacy implications of language models via contextual integrity theory,'' in \emph{The Twelfth International Conference on Learning Representations, {ICLR} 2024, Vienna, Austria, May 7-11, 2024}.\hskip 1em plus 0.5em minus 0.4em\relax OpenReview.net, 2024. [Online]. Available: \url{https://openreview.net/forum?id=gmg7t8b4s0}
\BIBentrySTDinterwordspacing

\bibitem{narouei2017identification}
M.~Narouei, H.~Khanpour, and H.~Takabi, ``Identification of access control policy sentences from natural language policy documents,'' in \emph{IFIP Annual Conference on Data and Applications Security and Privacy}.\hskip 1em plus 0.5em minus 0.4em\relax Springer, 2017, pp. 82--100.

\bibitem{nauman2010apex}
M.~Nauman, S.~Khan, and X.~Zhang, ``Apex: extending android permission model and enforcement with user-defined runtime constraints,'' in \emph{Proceedings of the 5th ACM symposium on information, computer and communications security}, 2010, pp. 328--332.

\bibitem{nguyen2015automatic}
T.-V.~T. Nguyen, N.~Fornara, and F.~Marfia, ``Automatic policy enforcement on semantic social data,'' \emph{Multiagent and Grid Systems}, vol.~11, no.~3, pp. 121--146, 2015.

\bibitem{10.1145/3508398.3511497}
M.~N. Nobi, R.~Krishnan, Y.~Huang, M.~Shakarami, and R.~Sandhu, ``Toward deep learning based access control,'' in \emph{Proceedings of the Twelfth ACM Conference on Data and Application Security and Privacy}, ser. CODASPY '22.\hskip 1em plus 0.5em minus 0.4em\relax New York, NY, USA: Association for Computing Machinery, 2022, p. 143–154.

\bibitem{norberg2007privacy}
P.~A. Norberg, D.~R. Horne, and D.~A. Horne, ``The privacy paradox: Personal information disclosure intentions versus behaviors,'' \emph{Journal of consumer affairs}, vol.~41, no.~1, pp. 100--126, 2007.

\bibitem{orenes2023using}
M.~Orenes-Vera, M.~Martonosi, and D.~Wentzlaff, ``Using llms to facilitate formal verification of rtl,'' \emph{arXiv preprint arXiv:2309.09437}, 2023.

\bibitem{planas2023modeling}
E.~Planas, S.~Mart{\'\i}nez, M.~Brambilla, and J.~Cabot, ``Modeling and enforcing access control policies in conversational user interfaces,'' \emph{Software and Systems Modeling}, vol.~22, no.~6, pp. 1925--1944, 2023.

\bibitem{rawte2023survey}
V.~Rawte, A.~Sheth, and A.~Das, ``A survey of hallucination in large foundation models,'' \emph{arXiv preprint arXiv:2309.05922}, 2023.

\bibitem{ren2023self}
J.~Ren, Y.~Zhao, T.~Vu, P.~J. Liu, and B.~Lakshminarayanan, ``Self-evaluation improves selective generation in large language models,'' in \emph{Proceedings on}.\hskip 1em plus 0.5em minus 0.4em\relax PMLR, 2023, pp. 49--64.

\bibitem{rossi2021seapp}
M.~Rossi, D.~Facchinetti, E.~Bacis, M.~Rosa, and S.~Paraboschi, ``$\{$SEApp$\}$: Bringing mandatory access control to android apps,'' in \emph{30th USENIX Security Symposium (USENIX Security 21)}, 2021, pp. 3613--3630.

\bibitem{scoccia2019empirical}
G.~L. Scoccia, A.~Peruma, V.~Pujols, B.~Christians, and D.~Krutz, ``An empirical history of permission requests and mistakes in open source android apps,'' in \emph{2019 IEEE/ACM 16th International Conference on Mining Software Repositories (MSR)}.\hskip 1em plus 0.5em minus 0.4em\relax IEEE, 2019, pp. 597--601.

\bibitem{scoccia2019permission}
G.~L. Scoccia, A.~Peruma, V.~Pujols, I.~Malavolta, and D.~E. Krutz, ``Permission issues in open-source android apps: An exploratory study,'' in \emph{2019 19th International Working Conference on Source Code Analysis and Manipulation (SCAM)}.\hskip 1em plus 0.5em minus 0.4em\relax IEEE, 2019, pp. 238--249.

\bibitem{shan2024automatic}
F.~Shan, Z.~Wang, M.~Liu, and M.~Zhang, ``Automatic generation of attribute-based access control policies from natural language documents.'' \emph{Computers, Materials \& Continua}, vol.~80, no.~3, 2024.

\bibitem{shi2011controlled}
L.~Shi and D.~W. Chadwick, ``A controlled natural language interface for authoring access control policies,'' in \emph{proceedings of the 2011 ACM Symposium on Applied Computing}, 2011, pp. 1524--1530.

\bibitem{smullen2020best}
D.~Smullen, Y.~Feng, S.~A. Zhang, and N.~Sadeh, ``The best of both worlds: Mitigating trade-offs between accuracy and user burden in capturing mobile app privacy preferences,'' \emph{Proceedings on Privacy Enhancing Technologies}, 2020.

\bibitem{song2024preference}
\BIBentryALTinterwordspacing
F.~Song, B.~Yu, M.~Li, H.~Yu, F.~Huang, Y.~Li, and H.~Wang, ``Preference ranking optimization for human alignment,'' in \emph{Thirty-Eighth {AAAI} Conference on Artificial Intelligence, {AAAI} 2024, Thirty-Sixth Conference on Innovative Applications of Artificial Intelligence, {IAAI} 2024, Fourteenth Symposium on Educational Advances in Artificial Intelligence, {EAAI} 2014, February 20-27, 2024, Vancouver, Canada}, M.~J. Wooldridge, J.~G. Dy, and S.~Natarajan, Eds.\hskip 1em plus 0.5em minus 0.4em\relax {AAAI} Press, 2024, pp. 18\,990--18\,998. [Online]. Available: \url{https://doi.org/10.1609/aaai.v38i17.29865}
\BIBentrySTDinterwordspacing

\bibitem{sonune2025lmn}
P.~Sonune, R.~Rai, S.~Sural, V.~Atluri, and A.~Kundu, ``Lmn: A tool for generating machine enforceable policies from natural language access control rules using llms,'' \emph{arXiv preprint arXiv:2502.12460}, 2025.

\bibitem{spencer1999flask}
R.~Spencer, S.~Smalley, P.~Loscocco, M.~Hibler, D.~Andersen, and J.~Lepreau, ``The flask security architecture: System support for diverse security policies,'' in \emph{8th USENIX Security Symposium (USENIX Security 99)}, 1999.

\bibitem{subramaniam2025deploiapplyingnl2sqlsynthesize}
\BIBentryALTinterwordspacing
P.~Subramaniam and S.~Krishnan, ``Deploi: Applying nl2sql to synthesize and audit database access control,'' 2025. [Online]. Available: \url{https://arxiv.org/abs/2402.07332}
\BIBentrySTDinterwordspacing

\bibitem{tjuatja2024llms}
L.~Tjuatja, V.~Chen, T.~Wu, A.~Talwalkwar, and G.~Neubig, ``Do llms exhibit human-like response biases? a case study in survey design,'' \emph{Transactions of the Association for Computational Linguistics}, vol.~12, pp. 1011--1026, 2024.

\bibitem{turtleguard}
\BIBentryALTinterwordspacing
L.~Tsai, P.~Wijesekera, J.~Reardon, I.~Reyes, S.~Egelman, D.~Wagner, N.~Good, and J.-W. Chen, ``Turtle guard: Helping android users apply contextual privacy preferences,'' in \emph{Thirteenth Symposium on Usable Privacy and Security (SOUPS 2017)}.\hskip 1em plus 0.5em minus 0.4em\relax Santa Clara, CA: USENIX Association, Jul. 2017, pp. 145--162. [Online]. Available: \url{https://www.usenix.org/conference/soups2017/technical-sessions/presentation/tsai}
\BIBentrySTDinterwordspacing

\bibitem{vance2019}
\BIBentryALTinterwordspacing
A.~Vance, D.~Eargle, J.~L. Jenkins, C.~B. Kirwan, and B.~B. Anderson, ``The fog of warnings: How non-essential notifications blur with security warnings,'' in \emph{Fifteenth Symposium on Usable Privacy and Security, {SOUPS} 2019, Santa Clara, CA, USA, August 11-13, 2019}, H.~R. Lipford, Ed.\hskip 1em plus 0.5em minus 0.4em\relax {USENIX} Association, 2019. [Online]. Available: \url{https://www.usenix.org/conference/soups2019/presentation/vance}
\BIBentrySTDinterwordspacing

\bibitem{vashurin2025benchmarking}
R.~Vashurin, E.~Fadeeva, A.~Vazhentsev, L.~Rvanova, D.~Vasilev, A.~Tsvigun, S.~Petrakov, R.~Xing, A.~Sadallah, K.~Grishchenkov \emph{et~al.}, ``Benchmarking uncertainty quantification methods for large language models with lm-polygraph,'' \emph{Transactions of the Association for Computational Linguistics}, vol.~13, pp. 220--248, 2025.

\bibitem{wash2010folk}
R.~Wash, ``Folk models of home computer security,'' in \emph{Proceedings of the Sixth Symposium on Usable Privacy and Security}, 2010, pp. 1--16.

\bibitem{wijesekera2015android}
P.~Wijesekera, A.~Baokar, A.~Hosseini, S.~Egelman, D.~Wagner, and K.~Beznosov, ``Android permissions remystified: A field study on contextual integrity,'' in \emph{24th USENIX Security Symposium (USENIX Security 15)}, 2015, pp. 499--514.

\bibitem{wijesekera2017feasibility}
P.~Wijesekera, A.~Baokar, L.~Tsai, J.~Reardon, S.~Egelman, D.~Wagner, and K.~Beznosov, ``The feasibility of dynamically granted permissions: Aligning mobile privacy with user preferences,'' in \emph{2017 IEEE Symposium on Security and Privacy (SP)}.\hskip 1em plus 0.5em minus 0.4em\relax IEEE, 2017, pp. 1077--1093.

\bibitem{wijesekera2018contextualizing}
P.~Wijesekera, J.~Reardon, I.~Reyes, L.~Tsai, J.-W. Chen, N.~Good, D.~Wagner, K.~Beznosov, and S.~Egelman, ``Contextualizing privacy decisions for better prediction (and protection),'' in \emph{Proceedings of the 2018 CHI Conference on Human Factors in Computing Systems}, 2018, pp. 1--13.

\bibitem{wu2025towards}
Y.~Wu, K.~Yang, F.~Roesner, T.~Kohno, N.~Zhang, and U.~Iqbal, ``Towards automating data access permissions in ai agents,'' \emph{arXiv preprint arXiv:2511.17959}, 2025.

\bibitem{xiao2020android}
J.~Xiao, S.~Chen, Q.~He, Z.~Feng, and X.~Xue, ``An android application risk evaluation framework based on minimum permission set identification,'' \emph{Journal of Systems and Software}, vol. 163, p. 110533, 2020.

\bibitem{xiao2012automated}
X.~Xiao, A.~Paradkar, S.~Thummalapenta, and T.~Xie, ``Automated extraction of security policies from natural-language software documents,'' in \emph{Proceedings of the ACM SIGSOFT 20th International Symposium on the Foundations of Software Engineering}, 2012, pp. 1--11.

\bibitem{yang2025extraction}
M.~Yang, V.~Atluri, S.~Sural, and A.~Kundu, ``Extraction of machine enforceable abac policies from natural language text using llm knowledge distillation,'' in \emph{Proceedings of the 30th ACM Symposium on Access Control Models and Technologies}, 2025, pp. 157--168.

\bibitem{zhang2015finedroid}
Y.~Zhang, M.~Yang, G.~Gu, and H.~Chen, ``Finedroid: Enforcing permissions with system-wide application execution context,'' in \emph{International Conference on Security and Privacy in Communication Systems}.\hskip 1em plus 0.5em minus 0.4em\relax Springer, 2015, pp. 3--22.

\bibitem{zheng2023judging}
L.~Zheng, W.-L. Chiang, Y.~Sheng, S.~Zhuang, Z.~Wu, Y.~Zhuang, Z.~Lin, Z.~Li, D.~Li, E.~Xing \emph{et~al.}, ``Judging llm-as-a-judge with mt-bench and chatbot arena,'' \emph{Advances in neural information processing systems}, vol.~36, pp. 46\,595--46\,623, 2023.

\end{thebibliography}

\appendices

\section{Ethical Considerations}
We evaluate the ethical implications of our work by identifying affected stakeholders and assessing the impacts of both the \textit{research process} (human-subject data collection and live LLM interaction) and the \textit{publication of results}. Further, the study was reviewed and approved by the authors' institutional ethics review board (IRB) prior to data collection.

\paragraph{Stakeholders}
We identified the following stakeholders:
\textit{Study participants} provide natural-language statements about permission preferences and make access control decisions in hypothetical mobile app scenarios. Their interests include informed consent, privacy of responses, protection from harmful or inappropriate content, and fair compensation.
The \textit{research team} relies on ethically collected data to evaluate whether LLMs can model personalized access control decisions.
Third-party infrastructure processes chatbot interactions. These \textit{LLM and cloud providers} are indirectly affected by expectations around data handling and responsible research use.
\textit{Future Designers, Users and Society} Future designers and users of automated access control systems—and society more broadly—may be affected by how such systems shape privacy, agency, and fairness in security-critical decisions.

\paragraph{Benefits}
This work provides empirical evidence on whether LLMs can capture user-specific access control preferences expressed in natural language. If successful, such systems may reduce the cognitive burden associated with repeated permission prompts and enable more user-aligned automation. Publishing these findings also informs the security community about the limits of LLM-based personalization, supporting more cautious and transparent system design.

\paragraph{Risks to Participants}
Participants may inadvertently disclose sensitive information during free-text interaction with a live chatbot or encounter unexpected or inappropriate model outputs. Even in the absence of tangible harm, such outcomes would violate participants' expectations of privacy or comfort.

\paragraph{Mitigations}
We implemented safeguards to reduce risks during the study:

\begin{itemize}[nosep]
    \item \textbf{Informed consent, withdrawal, and awareness.} Participants were admitted only after providing informed consent and could withdraw at any time; data from withdrawn participants were excluded from analysis without affecting compensation. Participants were explicitly informed about the use of a third-party LLM and instructed not to provide sensitive or identifying information. After completing the study, participants were debriefed and informed that some examples were hypothetical, and encouraged to critically evaluate outputs produced by automated systems.
    \item \textbf{Data minimization and protection.} Data was collected using pseudonymous identifiers; raw demographics are not published, and textual data is checked to remove identifying content.
    \item \textbf{Voluntary participation and fair compensation.} Participants were compensated at or above local minimum wage (USD 8.75 for an estimated 30-minute study, with an optional bonus for additional tasks) and could withdraw without penalty. Partial participation was remunerated with a proportional payout.
    \item \textbf{Responsible infrastructure use.} We used dedicated LLM deployments, approved by our institution, that do not retain data or use it for model training. Content filtering and careful prompting were applied to reduce harmful outputs.
\end{itemize}

\paragraph{Risks to Society}
Automating access control raises broader concerns about loss of user agency. There is also a risk that results are misread as endorsing autonomous decision-making, despite known limitations and failure modes.

\paragraph{Decision to Conduct and Publish}
We determined that conducting and publishing this study is ethically justified. The risks to participants are minimal and mitigated through consent, data protection, and infrastructure choices, while the benefits include improved understanding of how LLM-based automation may affect privacy-critical decisions. From a broader perspective, publishing these findings serves the public interest by enabling more informed and cautious deployment of LLMs in access control, rather than allowing such systems to be adopted without empirical scrutiny.

\section{Data quality}
\label{appx:dataquality}
To mitigate random answers, we implemented a \textit{commitment request} to provide thoughtful answers at the start of the data collection, as it was shown to successfully increase participants' involvement~\cite{qualtrics_commitment}, and two \textit{instruction manipulation} attention checks, where participants were clearly instructed to select a specific value.
In line with platform guidelines, we exclude participants failing the commitment check or both attention checks.

To ensure genuine answers to participants' \textit{privacy statements}, we (i) explicitly asked participants to refrain from using AI, and (ii) employed AI detection heuristics: participants with at least one copy-pasted answer, or where the keystroke count unnaturally matches exactly the length of their responses, are excluded from the dataset.
            
\section{Detailed Breakdown of Scenario Tasks} \label{appx:scenario_tasks}
Table~\ref{tab:scenario_table_expert_ordered} presents an overview of the scenario tasks, split by task type. 
It lists the total number of user decisions gathered, the outcome of the majority vote (either \texttt{allow} or \texttt{deny}), and the deny rate (ratio of \texttt{deny} decisions over all decisions made for a task). It further lists the decision made by the generic versions of GPT-4o and Mistral (i.e., $G_{\text{4o}}$ and $G_{\text{Mi}}$).
For the personalized $P_{\text{4o}}$ and $P_{\text{Mi}}$ models, the table lists their deny rate (i.e., the number of times the LLM decided \texttt{deny} for a user, divided by all decisions the LLM made for this scenario), and the agreement rate of the LLM decisions with the individual user decisions.

\clearpage
\begin{sidewaystable*}[p]
\centering
\caption{Scenario study tasks, divided by \discretion tasks, \essential tasks, and \critical tasks. For each scenario, we report the number of participants that made a decision for that task (\#), the result of the majority vote ($Maj.$), and the percentage of participants that made a \texttt{deny} decision ($D\%$). For GPT-4o and Mistral, we then report the decisions of the generic LLM, the deny rate of the personalized LLM, and the agreement of the personalized LLM with individual users' decisions.} 
\label{tab:scenario_table_expert_ordered}
{\setlength{\tabcolsep}{2pt}%
 \renewcommand{\arraystretch}{0.9}%
 \setlength{\aboverulesep}{0.2ex}\setlength{\belowrulesep}{0.2ex}%
 \scriptsize
\begin{tabular*}{\textheight}{@{\extracolsep{\fill}}llp{8.5cm}rlrrrrrrr@{}}
\toprule
\multicolumn{4}{c}{}  & \multicolumn{2}{c}{User} & \multicolumn{3}{c}{GPT-4o} & \multicolumn{3}{c}{Mistral} \\
\cmidrule(lr){5-6}\cmidrule(lr){7-9} \cmidrule(l){10-12}
App & Permission & Scenario & \# & Maj. & D\% & $G_{4o}$ & $P_{4o}$,D\% & $P_{4o}$,Agg\% & $G_{Mi}$ & $P_{Mi}$,D\% & $P_{Mi}$,Agg\% \\
\midrule
Airbnb & Calendar & The app offers you "Early Access", the opportunity to test new features that will be released in a future new app version. To continue, you need to grant the app access to your calendar.  & 138 & D & 59 & A & 53 & 55 & D & 91 & 59 \\
Amazon & Contacts & You want to send a digital gift card to a friend. You click on "Enter an email for each recipient". The app prompts you to grant access to your contacts in order to auto-fill the recipient's contact information. & 182 & D & 72 & A & 54 & 71 & D & 75 & 78 \\
Amazon & Location & The app offers you access to local deals and discounts. It requests location access to enable this. & 209 & A & 26 & O & 48 & 61 & A & 36 & 70 \\
Amazon & Photos & You browse the badges you have received in your account overview. The app offers you the "Added a profile picture badge", which you can get by adding a profile picture. The app requests access to your photos. & 194 & D & 69 & A & 67 & 75 & D & 75 & 73 \\
Duolingo & Contacts & You are about to start a lesson. The app asks for your permission to access your contacts in exchange for 3 days of "Super" subscription. & 138 & D & 86 & D & 99 & 86 & D & 100 & 86 \\
Duolingo & Contacts & You set up the app. It asks to access your contacts to find friends using the app for the best experience.  & 122 & D & 80 & D & 78 & 77 & D & 85 & 78 \\
Facebook & Contacts & You set up the app. It asks for access to your contacts to find your friends faster. & 177 & D & 64 & D & 85 & 66 & D & 94 & 65 \\
Gmail & Calendar & You want to set up a vacation responder. The app offers to set up vacation responders automatically in the future. It requests access to the calendar.  & 179 & A & 31 & A & 37 & 66 & A & 52 & 55 \\
Netflix & Microphone & You click play to start watching a series. The app offers to adjust the volume based on noise around you and requests access to the microphone. & 191 & D & 61 & O & 19 & 47 & O & 47 & 50 \\
Tinder & Calendar & You set up the app. After answering questions about your habits, it requests access to your calendar to find potential matches that attend the same events as you. & 90 & D & 67 & D & 82 & 71 & D & 93 & 67 \\
Tinder & Contacts & You set up the app. It requests access to your contacts to hide your profile from people you know. & 103 & D & 55 & A & 56 & 62 & D & 67 & 61 \\
Tinder & Photos & You set up the app. It requests access to your photos so that you can add pictures to your profile.  & 90 & A & 16 & A & 11 & 80 & A & 27 & 73 \\
Uber & Calendar & You set up the app. It requests access to your calendar to suggest ride options based on calendar events with addresses specified. & 152 & D & 74 & A & 61 & 60 & A & 61 & 56 \\
Uber & Camera & You set up the app. After entering your payment info, the app requests camera access to take a picture of you which drivers can use to identify you. & 169 & A & 42 & A & 7 & 60 & O & 2 & 59 \\
Whatsapp & Contacts & You set up the app. It requests access to your contacts to find people who are also using the app. & 174 & A & 17 & A & 33 & 65 & A & 49 & 58 \\
\midrule
Booking & Location & You are at home and want to search for accommodations for friends planning to visit you. You click "Enter your destination" and then "Around current location". The app asks for access to your location to show you nearby options. & 157 & A & 12 & O & 1 & 88 & O & 0 & 88 \\
Chatgpt & Microphone & You want to start a conversation and press "[microphone]". The app request access to the microphone.  & 175 & A & 8 & A & 0 & 92 & A & 0 & 92 \\
Snapchat & Camera & The app asks for permission to access the camera to be able to make snaps. & 125 & A & 4 & A & 1 & 94 & A & 2 & 94 \\
Tinder & Location & You set up the app. It asks for your location to be able to continue. & 97 & A & 16 & A & 9 & 78 & A & 28 & 74 \\
Uber & Microphone & You want to call your diver through the app and press "[microphone]".  & 179 & A & 8 & O & 0 & 92 & O & 0 & 91 \\
Whatsapp & Camera & You want to start a video call and press "[camera]". The app requests access to your camera.  & 188 & A & 2 & A & 0 & 98 & A & 0 & 97 \\
\midrule
Airbnb & Location & You want to search for accommodations for a trip you plan to do in three months. You click "Start your search". The app asks for access to your location. & 147 & A & 29 & D & 88 & 33 & O & 11 & 70 \\
Facebook & Photos & You set up the app. It asks for access to your photos to pre-index your photo library, allowing for a more efficient picture upload later on.  & 170 & D & 60 & D & 81 & 60 & D & 83 & 63 \\
Instagram & Camera & You like a post. The app requests camera access. & 178 & D & 64 & D & 100 & 64 & D & 99 & 64 \\
Snapchat & Microphone & You set up the app. It asks you to enable the permission to make sign up easy. & 138 & A & 45 & D & 98 & 48 & D & 88 & 48 \\
Spotify & Location & You open the app and start browsing for music to listen to. The app asks for your location. & 186 & D & 63 & D & 91 & 66 & D & 89 & 64 \\
Whatsapp & Photos & The app warns you that your storage is full and requests access to your photos to delete blurry pictures. & 162 & D & 62 & A & 60 & 57 & D & 76 & 56 \\
\bottomrule
\end{tabular*}}

\end{sidewaystable*}

\clearpage

\end{document}